\documentclass[11pt,letterpaper]{article}
\usepackage[T1]{fontenc}
\usepackage[english]{babel}
\usepackage{times}
\usepackage[letterpaper,margin=1in]{geometry}
\usepackage{amsmath,amstext,amsbsy,amsopn,amsthm,amsfonts,amssymb}
\usepackage{booktabs,array,colortbl}
\usepackage[]{graphicx}
\usepackage{rotate}
\usepackage{wrapfig}
\usepackage{caption}
\usepackage{subfigure}
\usepackage{lastpage}

\usepackage[small]{titlesec}
\usepackage{paralist}
\usepackage{verbatim} % For comment
\usepackage[usenames,dvipsnames,svgnames]{xcolor}
\usepackage{tikz} %% for GANTT chart
\usepackage{pgfgantt}
\usepackage{pgfcalendar}
\usetikzlibrary{shadows}
\usetikzlibrary{shadings}
\usepackage{multirow}
\usepackage{algorithmic}
\usepackage{algorithm}
\usepackage{fancyhdr}
\providecommand{\keywords}[1]{\textbf{\textit{Keywords---}} #1}
\renewenvironment{thebibliography}[1]{%
	\begin{oldthebibliography}{#1}%
		\setlength{\parskip}{0.0cm}%
		\setlength{\itemsep}{0.0cm}%
	}%
	{%
	\end{oldthebibliography}%
}

\begin{document}

%\conferenceinfo{DAC'12,} {June 2--6, 2013, Austin, USA.}
%\CopyrightYear{2013}
%\crdata{XXXXXXXXXX}
%\clubpenalty=10000
%\widowpenalty = 10000

%\title{iVAMS: Intelligent Metamodel-Integrated Verilog-AMS for Circuit-Accurate System-Level Mixed-Signal Design Exploration}

%\title{iVAMS: Machine-Learning-Metamodel Integrated Verilog-AMS for Fast Analog Block Optimization}

\title{\textbf{iVAMS 2.0}: Machine-Learning-Metamodel-Integrated Intelligent Verilog-AMS for Fast and Accurate Mixed-Signal Design Optimization}

%\numberofauthors{1}
%\author{
%\alignauthor Geng Zheng$^1$, Saraju P. Mohanty$^2$, Elias Kougianos$^3$\\
%\affaddr{NanoSystem Design Laboratory (NSDL)$^{1,2,3}$}\\
%\affaddr{Dept.of Computer Science and Engineering$^{1,2}$
%and Dept. of Electrical Engineering Technology$^3$}\\
%\affaddr{Department of Electrical Engineering Technology$^3$}\\
%\affaddr{University of North Texas  Denton, TX 76203.$^{1,2,3,4}$}\\
%\email{gengzheng@my.unt.edu$^1$, saraju.mohanty@unt.edu$^2$, eliask@unt.edu$^3$}}

\maketitle

%\author{
%Saraju P. Mohanty \\
%	Department of Computer Science and Engineering \\
%University of North Texas\\
%	 Denton, TX 76203 \\
%	\texttt{saraju.mohanty@unt.edu} \\
	%% examples of more authors
%\and
%	Elias Kougianos \\
%	Department of Electrical Engineering\\
%University of North Texas\\
% Denton, TX 76203 \\
%	\texttt{elias.kougianos@unt.edu} \\
%}

\author{
\begin{center}
	\begin{tabular}{cc}
		Saraju P. Mohanty & Elias Kougianos \\
		Computer Science and Engineering & Engineering Technology \\
		University of North Texas, Denton, TX 76203. & University of North Texas, Denton, TX 76203. \\
Email: \texttt{saraju.mohanty@unt.edu} & Email: \texttt{elias.kougianos@unt.edu}
	\end{tabular}
\end{center}
}

\cfoot{Page -- \thepage-of-\pageref{LastPage}}



\begin{abstract}
The gap between abstraction levels in analog design is a major obstacle for advancing analog and mixed-signal (AMS) design automation and computer-aided design (CAD). Intelligent models for low-level analog building blocks are needed to bridge the accuracy gap between behavioral and transistor-level simulations. The models should be able to accurately estimate the characteristics of the analog block over a large design space. Machine learning (ML) models based on actual silicon have the capabilities of capturing detailed characteristics of complex designs. In this paper, a ML model called Artificial Neural Network Metamodels (ANNM) have been explored to capture the highly nonlinear nature of analog blocks. The application of these intelligent models to multi-objective analog block optimization is demonstrated. Parameterized behavioral models in Verilog-AMS based on the artificial neural network metamodels are constructed for efficient system-level design exploration. To the best of the authors' knowledge \emph{this is the first paper to integrate artificial neural network models in Verilog-AMS, which is called iVAMS 2.0}. To demonstrate the application of iVAMS 2.0, this paper presents two case studies: an operational amplifier (OP-AMP) and a phase-locked loop (PLL). A biologically-inspired ``firefly optimization algorithm'' is applied to an OP-AMP design in the iVAMS 2.0 framework. The optimization process is sped up by 5580$\times$ due to the use of iVAMS with negligible loss in accuracy. Similarly, for a PLL design, the physical design aware ANNs are trained and used as metamodels to predict its frequency, locking time, and power. Thorough experimental results demonstrate that only 100 sample points are sufficient for ANNs to predict the output of circuits with 21 design parameters within 3\% accuracy, which improves the accuracy by 56\% as compared to polynomial metamodels. A proposed artificial bee colony (ABC) based algorithm performs optimization over the ANN metamodels of the PLL. It is observed that the ANN metamodels achieve more accurate results than polynomial metamodels with shorter optimization time.
\end{abstract}

%%%%%%%%%%%%%%%%%%%%%%%%%%%%%%%%%%%%%%%%%%%%%%%%%%%%%%%%%%%%%%%%%%%%%%%%%%%%%%%
%\category{B.7.1}{Integrated Circuits}{Types and Design Styles---}
%{VLSI (very large scale integration)}

%\terms{Design, Optimization}

\keywords{Machine Learning Models, Artificial Neural Networks (ANN), Intelligent Verilog-AMS, Metamodeling, System Simulation, Mixed-Signal Design, Behavioral Simulation, Verilog-AMS Modeling, OP-AMP, Phase-Locked Loop (PLL)}

%%%%%%%%%%%%%%%%%%%%%%%%%%%%%%%%%%%%%%%%%%%%%%%
\section{Introduction}

% Introduction and motivation.

% 1) iVAMS
% 2) Op Amp
% \begin{figure}[!ht]
% \centerline{
% \includegraphics[scale=1]{figs/sch2lay}}
% \caption{Design iterations from baseline schematic design to optimal layout design.}
% \label{fig_sch2lay}
% \end{figure}

Competitive time-to-market and design constraints discourage or prohibit the use of slow exhaustive design space exploration to reach fully optimal performance for complex nano-CMOS mixed-signal designs. 
The trend of integrated circuit (IC) development toward increasing levels of integration has never stopped. Design automation tools and flows for digital IC systems have followed this trend due to the fact that digital circuits have higher regularity and high margin for noise and process variations \cite{Mohanty_Book_2015_Mixed-Signal,mohanty2013incorporating}. Thus their performance can be well controlled throughout different abstraction levels from top to bottom. In contrast, a hierarchical way to tightly control and predict the performance of analog and mixed-signal (AMS) blocks at every design phase has not yet been established in industrial practice.

Modern AMS systems are complex and contain numerous nanometer devices. In order to achieve high performance and high yield, a system must be optimized at both system and circuit levels \cite{Buffeteau_PATMOS_2018,mohanty_patent_2015_intelligent,mohanty2012statistical}.
For a top-down design approach, this process starts with designing and optimizing the system with sub-block models at high levels of abstraction. The specifications for each sub-block that lead to the best system performance are then obtained. Each sub-block is then designed and optimized toward these specifications. The issue with this approach is that generating a perfect model, if it is even possible, takes significant effort. Therefore some characteristics of the sub-blocks are ignored at the high level simulation. This makes the obtained sub-block specification compliance less reliable. In other words, optimizing the sub-blocks using these specifications does not guarantee satisfying system performance \cite{mohanty_patent_2015_methodology}. Usually numerous design iterations are required. To make the problem worse, performing time-domain simulations (transient analyses) is an important step when evaluating an AMS system and this process consumes large amount of computational time. The parasitics of the physical design of the sub-blocks further aggravate this problem considering the fact that redoing the layout manually may be necessary.

\begin{figure}[htbp]
\centering
\includegraphics[width=0.65\textwidth]{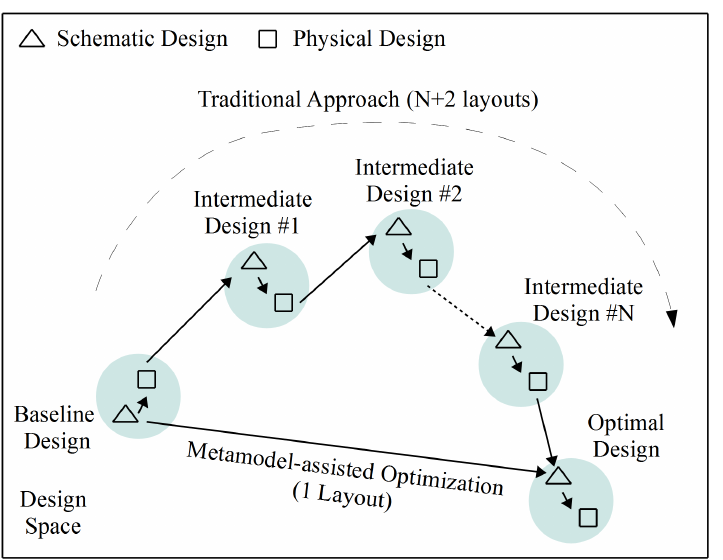}
\caption{Design iterations from baseline schematic design to optimal layout design.}
\label{fig_sch2lay}
\end{figure}

This work proposes developing a design and optimization methodology that not only can generate compliant designs but can also complete the whole process in much shorter time, compared to conventional approaches. Fig.~\ref{fig_sch2lay} illustrates the potential benefit of using a metamodel-assisted optimization flow compared to the traditional flow \cite{Mohanty_Book_2015_Mixed-Signal,mohanty_patent_2015_methodology,Mohanty_Invited-Talk_2012}. In the traditional approach, the design time for each iteration is prolonged by generating, extracting and simulating the physical design. While the metamodel-assisted approach provides a direct path from the baseline schematic design to the optimized schematic design whose corresponding physical design performance is expected to equal or better the performance of the design obtained from the traditional approach.

In a hierarchical approach, design information should propagate seamlessly between abstraction levels. A critical part of this approach is efficient and accurate surrogate models for low-level AMS building blocks \cite{Kotti_SMACD_2018}. When a design variable value, such as the size of a transistor, changes in the building block, this model should capture the resultant changes of the block characteristics immediately and pass them to higher levels. This model should also allow a system-level algorithm to fine tune its local design variable values to optimize the overall system performance. We propose \textbf{the following requirements} for such a model for blocks of mixed-signal systems \cite{Zheng_ASAP2013,Zheng_DAC2013}:
\begin{enumerate}
%\begin{inparaenum}[\itshape a\upshape)]
\item
The model should be capable of modeling the building block performance metrics for fast design optimization;
\item
The model should be able to be used in high-level AMS behavioral simulations;
\item
The model should be parameterized so that it can capture the entire response surface of the building block with reasonable accuracy over a large design space;
\item
The construction and use of such models should only cost a small portion of an analog designer's time and the CPU time required for this process should be moderate. %\end{inparaenum}
\end{enumerate}

The last requirement reflects the fact that the designer's time is more valuable than CPU time \cite{kra_2001}. While our aim is to minimize the CPU time, minimizing the burden imposed on the designer has higher priority. With the aforementioned considerations, we propose an intelligent metamodel integrated Verilog-AMS (iVAMS) module for analog blocks. iVAMS aims at closing the gap between abstraction levels in analog design which is currently regarded as the ``number one'' requirement for advancing AMS design automation \cite{gra_2012}. We introduced iVAMS 1.0 \cite{Mohanty_ArXiv_2019_iVAMS1} that integrated simple polynomial metamodels to accurate and fast optimization. This article discusses iVAMS 2.0 which integrates machine learning (ML) metamodels in Verilog-AMS for fast and accurate design optimization of large AMS-SoCs.

The rest of this paper is organized as follows: Section~\ref{sec_cont} lists the contributions of this paper. Section~\ref{sec_ivams} presents the concept of iVAMS.  Section~\ref{sec_prior} discusses related research. Section~\ref{sec_case_OP-AMP} uses a 90nm OP-AMP as a case study to demonstrate the generation and use of iVAMS for an analog block. Section~\ref{sec_prior} discusses related research. Section~\ref{sec_case_PLL} uses a 180nm PLL as a case study to demonstrate the generation and use of iVAMS for an AMS block. Section~\ref{sec_conc} concludes this paper and discusses future research.

%%%%%%%%%%%%%%%%%%%%%%%%%%%%%%%%%%%%%%%%%%%
\section{Contributions of this Paper}
\label{sec_cont}

The \textbf{overall contribution of this paper} is to enable system-level or behavioral modeling with circuit-level intelligent artificial neural network (ANN) metamodels such that the gap between system-level speed and circuit-level accuracy is bridged.

The \textbf{novel contributions of this paper} to the state-of-the-art can be summarized as follows:
\begin{enumerate}
%[\IEEEsetlabelwidth{12}]

\item
The concept of \textbf{i}ntelligent-metamodel integrated \textbf{V}erilog-\textbf{AMS} (\textbf{iVAMS});

\item
Schematic design of a 90nm CMS based OP-AMP;

\item
Intelligent metamodel generation with neural networks;

\item
A biologically-inspired firefly based algorithm for multi-objective OP-AMP optimization using iVAMS;

\item
Construction of a neural network integrated parameterized OP-AMP behavioral model in Verilog-AMS.

\item The paper combines an artificial bee colony (ABC) optimization algorithm (which is a swarm intelligence type of evolutionary computation) and ANN based metamodeling for fast and accurate nano-CMOS mixed-signal design exploration.

\item A non-polynomial metamodel based design optimization flow for analog/mixed-signal circuits is presented. For non-polynomial metamodeling, different architectures of ANNs are considered for trade-off analysis between speed and accuracy.

\item  As a practical demonstration of the use of the non-polynomial ANN metamodels, a physical design optimization of a 180 nm CMOS PLL is undertaken.

\item A biologically inspired tool, the ABC algorithm is used for optimization of the PLL physical design that uses the metamodels instead of the actual circuit (i.e. the parasitic aware netlist).

\item It is demonstrated that the ANN metamodel assisted optimization is faster and more accurate compared to the polynomial metamodel.
\end{enumerate}

Metamodeling based design flows are investigated as approaches to reduce design cycle time \cite{mohanty_patent_2015_intelligent,Okobiah_TVLSI_2013-Apr}. In this paper, the proposed non-polynomial metamodeling design flow speeds up the design process by creating accurate metamodels and uses them for optimization as surrogates of the actual design models. Metamodeling is used in a variety of different fields to predict the values of time consuming or expensive processes \cite{Barton2009WSC}. Creation of the metamodel starts by sampling the design space and then using mathematical approaches to create formula(e) for output prediction. For circuits, the simulation is performed using analog circuit simulators such as SPICE. There are different approaches to generating the predictive formula(e). Polynomial least square regression is the most common and widely used \cite{gar_2012}. Its simplicity is very attractive, but it is not efficient for very high dimensional circuits (many parameters) due to the number of required coefficients. To improve polynomial regression models, spline based functions can be used. But they also have the same limitations as regular polynomials for high dimensional data sets.

%%%%%%%%%%%%%%%%%%%%%%%%%%%%%%%%%%%%%%%%%%%%%%%%%%%%%%
\section{Proposed Intelligent Verilog-AMS (iVAMS)}
\label{sec_ivams}

\subsection{The iVAMS Concept}

\begin{figure}[t]
\centering
\includegraphics[width=0.70\textwidth]{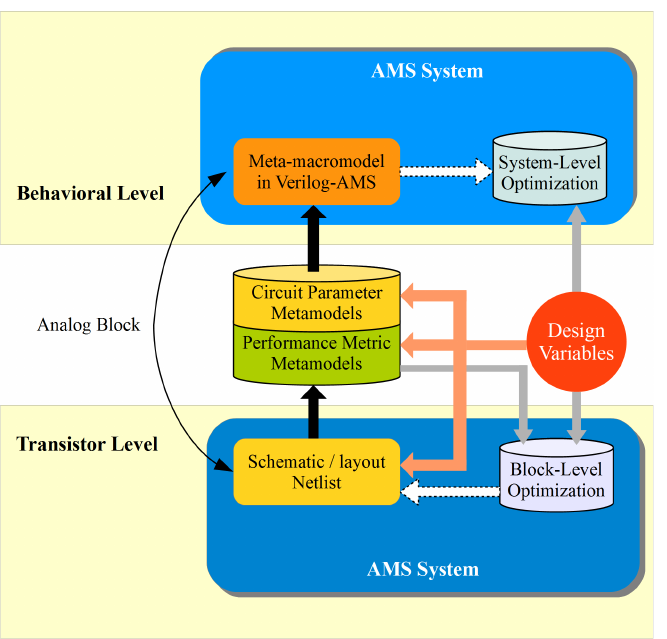}
\caption{The concept of the proposed iVAMS \cite{Mohanty_ArXiv_2019_iVAMS1}.}
\label{fig_ivams}
\end{figure}

Simulation is an essential part of design verification and optimization. This abstraction level is critical since it is the immediate level above the transistor-level netlist and thus determines the accuracy and runtime of the system verification/optimization. In the simulation process, parameterized behavioral models for the analog blocks are necessary.
Simulating an entire AMS system with transistor-level netlists, though very accurate, is formidable. A common solution is to simulate the AMS system at the behavioral level. Parameterized behavioral models share the same design variables as the transistor-level design. When the values of the design variables change, the values of the circuit parameters for the behavioral model change accordingly. Thus the impact of the transistor-level design changes on the behavioral level are captured. iVAMS provides such Verilog-AMS modules, as shown in Fig. \ref{fig_ivams} \cite{Mohanty_ArXiv_2019_iVAMS1}.

iVAMS is based on a set of artificial neural network (ANN) metamodels. The metamodels are sample-based  and fall into two categories: the performance metric metamodel (PMM) and the circuit parameter metamodel (CPM). Given an analog block, PMMs estimate its performance for any point in the design space. No actual circuit simulation is required in this estimation process. Thus, PMMs provide an ultra-fast way to explore the design space of the analog block. CPMs, on the other hand, estimate the circuit parameters required to construct a macromodel (\emph{not} a metamodel) for the analog block. By integrating the CPMs into a macromodel and describing them in Verilog-AMS, a parameterized behavioral model, called \textbf{meta-macromodel}, can be obtained. A macromodel is usually a white-box or grey-box model \cite{rom_2008} that retains certain physical information of the analog block's structure \cite{gom_1995,wang_2007,zha_2011}. A metamodel is a black-box model that describes the analog block using mathematical algorithms, without explicitly carrying any physical information about it \cite{Garitselov_TSM_2012-Feb,fang_2006}. Constructing a macromodel for an analog block requires physical insight and consumes a great portion of the designer's time. Constructing a metamodel requires sampling the design space, which consumes CPU time. Meta-macromodeling selects a suitable macromodel and estimates the required parameters using metamodels. This way a good trade-off is reached. The meta-macromodel is the core component for exploring the design space of the AMS system. The best (may be the most accurate or the fastest, depending on the requirement) ANN metamodels need to be selected for optimization. As a specific objective and constraint optimization, the PLL circuit is characterized for output frequency, power, and locking time. Similarly other characteristics can be chosen for other circuits such as operational amplifiers (OP-AMP) and analog-to-digital converters (ADC). A separate metamodel is created for each Figure-of-Merit (FoM) from the same sample set. Each single simulation calculates all FoMs so the number of simulations that are needed does not depend on the number of the metamodels that need to be created.

Artificial Neural Networks (ANNs) provide an alternative for creating very high dimensional models \cite{Wang2005AMC}. For a limited amount of simulations, the ANN preforms almost equally well for any number of parameters. Multi-layer networks are trained in parallel for every input by adjusting the corresponding weights for non-linear and linear functions. Once the network learns and conducts final adjustments for weights of the internal functions, it is able to predict the values with only the number of parameters times the number of layer functions in the model. This makes ANNs very robust. Finding the right architecture for a neural network usually requires some experimentation. This work analyzes different ANN structures and internal functions and applies generated metamodels for further multi-objective optimization algorithms to receive the optimal parameter values. The data to create the ANN are directly generated by SPICE. Once the ANN is generated, it can predict the output value very fast, due to its low complexity and simplicity. Hence, the prediction is orders of magnitude faster than SPICE.

\subsection{iVAMS 2.0 Generation}
\label{sec_gen}

\begin{figure*}[t]
\centering
\includegraphics[width=0.77\textwidth]{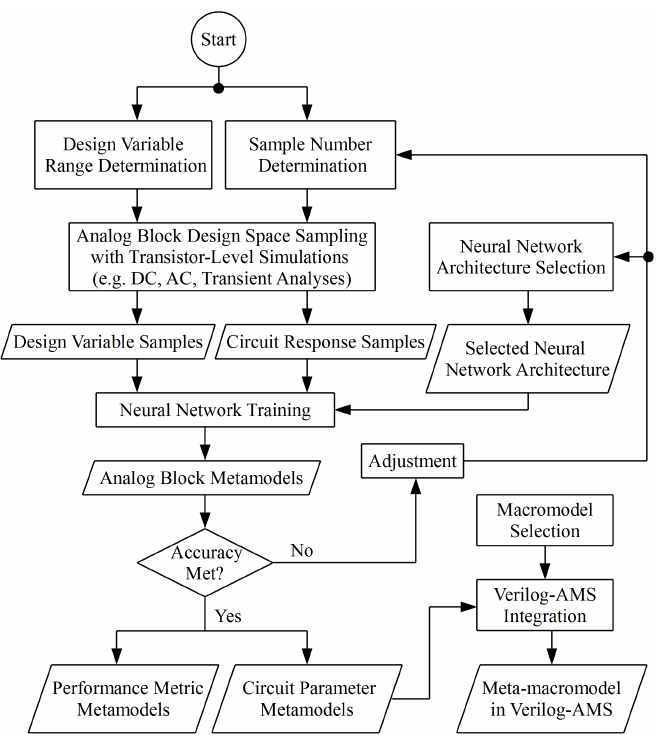}
\caption{The detailed steps of iVAMS 2.0 generation \cite{Zheng_ASAP2013}.}
\label{fig_flow}
\end{figure*}

Given an analog block, the goal of iVAMS generation is to obtain the artificial neural network metamodels (PMMs and CPMs) and the meta-macromodel integrated Verilog-AMS module. The iVAMS generation flow is shown in Fig.~\ref{fig_flow}. The process starts with determining the design variable ranges and the number of simulation samples. By sampling the analog block design space with transistor-level simulations, a number of design variable samples and samples of the circuit's responses of interest are obtained. Using these samples, neural networks with the selected architecture are then trained to accurately model the circuit responses. If the accuracy is insufficient, the designer can adjust the neural network architecture and/or add additional samples.

Once the artificial neural network metamodels with sufficient accuracy are obtained, the generated PMMs can be used in optimizing the analog block. In order to generate a meta-macromodel for the analog block in Verilog-AMS, a macromodel architecture is selected and implemented in Verilog-AMS. The computation of the circuit parameters for the macromodel is done using CPMs and is embedded in the \verb+initial+ block of the Verilog-AMS module.

\begin{figure}[htbp]
	\centering
	\includegraphics[width=0.67\textwidth]{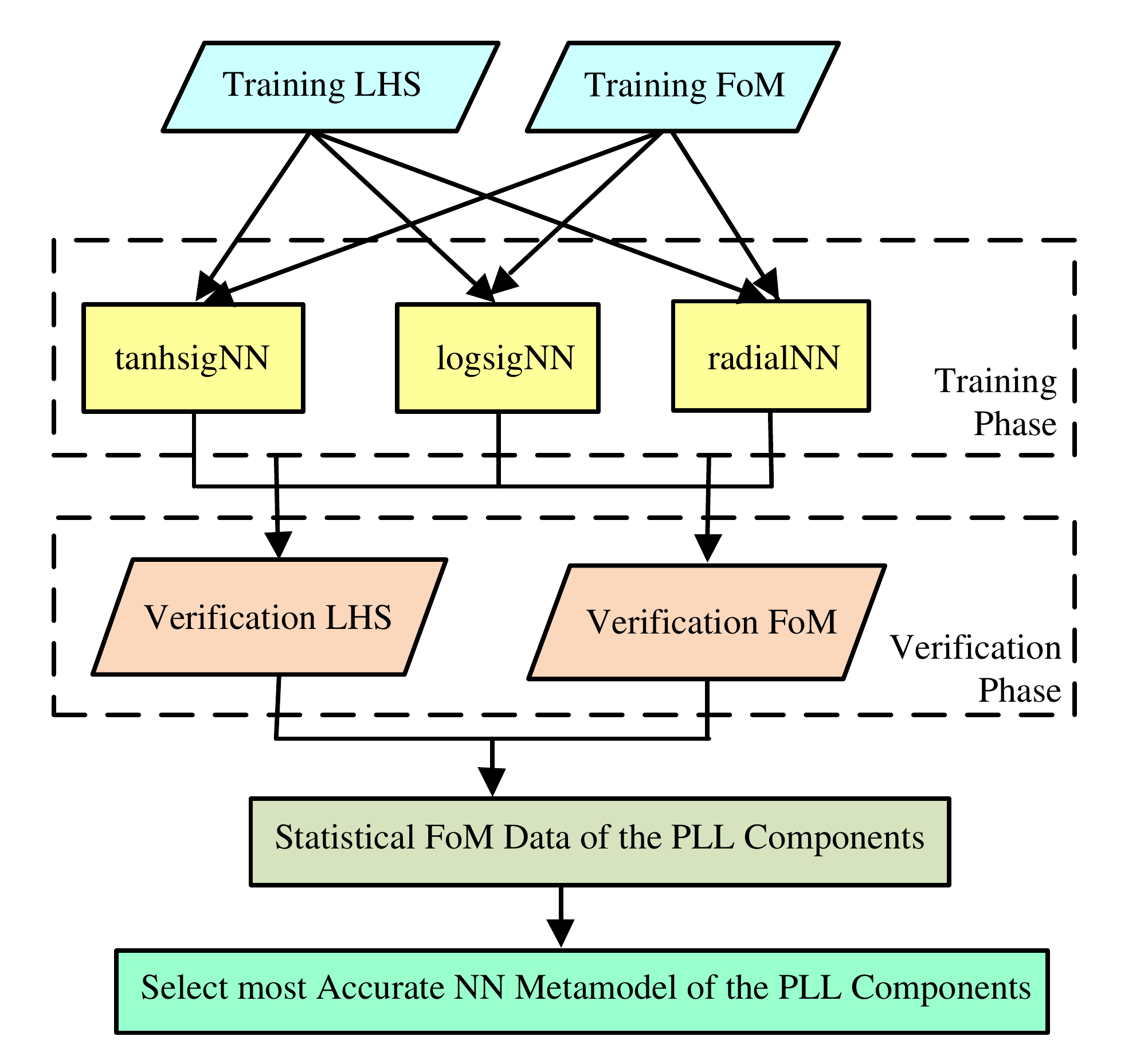}
	\caption{Flow for generating machine learning (artificial neural network or ANN) metamodel.}
	\label{fig:ANN_flow}
\end{figure}

Artificial neural network models are composed of a mass of fairly simple computational elements and rich interconnections between the elements. The ANN generation flow is shown in Figure \ref{fig:ANN_flow}. A training data set is generated using Latin Hyper-Cube Sampling (LHS) and the results are used for ANN training. The verification is conducted using a verification data set that is roughly 30\% the size of the training data set and is generated also using LHS. It is important that the training and verification data sets are distinct. Otherwise the performance of the ANN will be artificially inflated. Statistical fitting data is collected from both training and verification data. The best configuration metamodel is selected based on the statistical data, as discussed further in this Section.

ANNs operate in a parallel and distributed fashion which may resemble biological neural networks. Most ANNs have some sort of ``training'' rule by which the weights of the connections are adjusted on the basis of presented patterns. They normally have great potential for parallelism, since the computations of the components are independent of each other. It has been proven in the universal approximation theorem \cite{NIPS2017_7203} that a neural network with one hidden layer can estimate any continuous function that maps to real numbers.

\subsubsection{Multilayer Artificial Neural Networks (ANNs)}

\begin{figure}[t]
	\centering
	\includegraphics[width=0.67\textwidth]{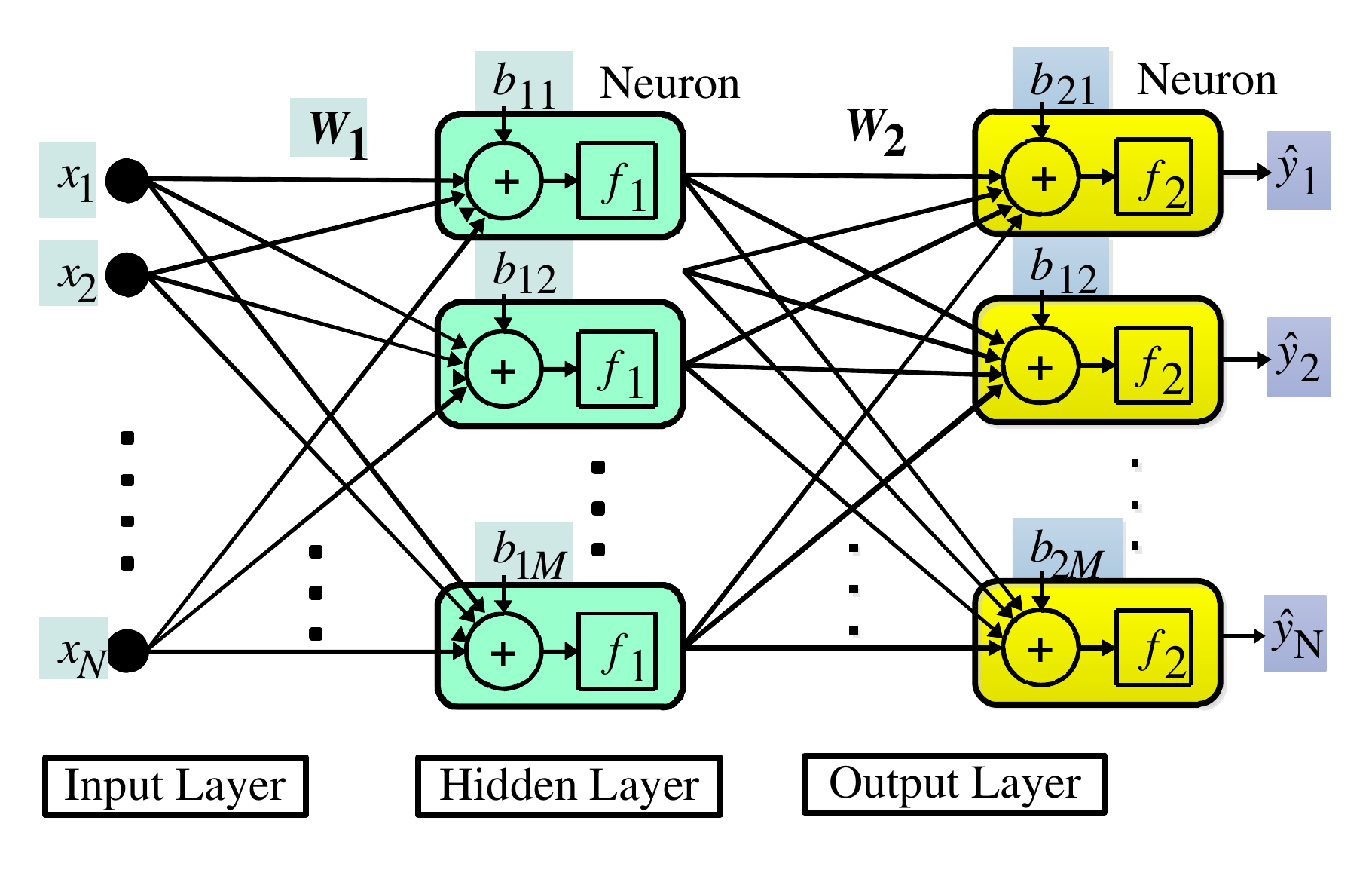}
	\caption{Feed forward ANN architecture. $N$ input parameters $x$ using weights for the interface between the hidden ($W_1$) and local layers ($W_2$) functions to produce the output responses $\hat{y}_j$. }
	\label{fig:ANN_Architecture}
\end{figure}

A multiple layer ANN consists of an input, a with nonlinear activation function in hidden layer, and linear activation function in the output layer (Fig. \ref{fig:ANN_Architecture}). Multilayer networks are very flexible and powerful due to their ability to represent nonlinear as well as linear functions. The multilayer ANN needs to have at least one non-linear function, otherwise a composition of linear functions becomes just another linear function.

The two common nonlinear activation functions that are usually used for the hidden layer are \cite{Mohanty_Book_2015_Mixed-Signal}:
\begin{equation}
b_j \left( v_j \right) = \left( \frac{1}{1+e^{-\lambda v_j}} \right), \text{ or}
\end{equation}
\begin{equation}
b_j \left( v_j \right)=\tanh(\lambda v_j),
\end{equation}
where $j$ denotes a neuron in the hidden layer, $b_j$ and $v_j$ are its input and output, respectively, and $\lambda$ is the neuron transfer function steepness.
The predicted output is given by:
\begin{equation}
\hat{y}= \sum_{j=1}^{d}{\beta_jb_j(v_j)+\beta_0},
\end{equation}
where $\beta_j$  is the weight in the output due to neuron $j$, $d$ is the number of neurons in the hidden layer and $\beta_0$ a constant.
On the other hand, a linear layer function has the following format:
\begin{equation}
v_i = \sum_{i=1}^{s}{w_{ji}x_i+w_{j0}},
\end{equation}
where $w_{ji}$ is the weight connection between the $j$th component in the hidden layer and the $i$th component of the input and $w_{j0}$ is a constant bias.
The ANN training is performed to minimize the least square criterion:
\begin{equation}
E=\sum_{k=1}^{n}{(y_k-\hat{y_k})^2},
\end{equation}
where $y_k$ and $\hat{y_k}$ are the actual and predicted responses, respectively, at the $k$-th training point (of $n$).

\subsubsection{Radial Neural Networks}

Radial neural networks are also two-layer networks. The first layer has radial base neurons, and calculates its weighted inputs with distance and its net input with a radial function. The second layer has linear neurons, and calculates its weighted input with a dot product function and its net inputs  by combining its weighted inputs and biases. Both layers have biases. The radial network mathematical model is as follows:
\begin{equation}
y=\sum_{i=1}^{N}{a_i \rho \left( \parallel x - c_i \parallel \right)},
\end{equation}
where $c_i$ is the center vector of neuron $i$, $x$ is the prediction point, $\rho$ is the neuron's transfer function and $a_i$ are the weighs of the linear neuron. $\parallel x - c_i \parallel$ is the Euclidean distance between $x$ and $c_i$. 

Initially the radial basis layer has no neurons. The following steps are repeated until the network's mean squared error falls below the desired goal. The network is simulated. The input vector with the greatest error is found. A radial basis neuron is added with weights equal to that vector. The linear layer weights are then recalculated to minimize the error.

Each neuron in the radial basis layer will output a value according to how close the input vector is to each neuron's weight vector. Thus, radial basis neurons with weight vectors quite different from the input vector have will have outputs near to zero. These small outputs have only a negligible effect on the linear output neurons.

In contrast, a radial basis neuron with a weight vector close to the input vector produces a value near 1. If a neuron has an output of 1 its output weights in the second layer pass their values to the linear neurons in the second layer.

\subsection{Standardizing or Normalizing Data during ML model generation}

The data set is generated from the RLCK parasitic aware netlist simulations. It include parasitic resistance (R), parasitic capacitances (C), parasitic self-inductances (L) as well as coupling inductances (K). The input data set is the same for every metamodel and is generated using LHS. LHS supports any amount of planes and is proven to work better than Monte Carlo due to the more even distribution of points while still incorporating the random factor that helps to detect nonlinearity. LHS divides each plane (parameter) into Latin squares and randomly picks a point from each square. An output is generated for each run from SPICE simulations saving each needed value to its own data set. Hence, each metamodel has its own target data set. This paper targets ANNs that have a single output with multiple inputs.

The validation and test data must be standardized or normalized using the statistics computed from the training data. It is desirable to either normalize or standardize the input data as the input dataset typically has a large dynamic range. If not, the training of higher values can outweigh the lower values and the ANN will not train properly. In this paper, 2 commonly used methods are applied to standardize the data:
\begin{enumerate}
	\item
	Normalizing to mean $\left( \mu \right)$ 0 and standard deviation $\left( \sigma \right)$ 1.
	
	\item
	Standardizing to midrange 0 and range 2 (from -1 to 1).
\end{enumerate}

\subsection{ANN Metamodel Selection Criteria}

There may be numerous metamodels created from the same sampled set. The Root Mean Square Error (RMSE) and correlation coefficient $R^2$ are common metrics used for goodness of fit. The RMSE is derived from the sum of square errors (SSE):
\begin{eqnarray}
RMSE & = & \sqrt{\frac{1}{N}SSE} \\
& = & \sqrt{\frac{1}{N} \sum_{k=1}^{N}{(y({x}_{k})-\hat{y} ({x }_{k }))}^{2}},
\label{RMSE}
\end{eqnarray}
where $y$ are the actual simulation result values and $\hat{y}$ are the results of the metamodel at the same location as the simulation point. $R^2$ is the coefficient of determination, which predicts the probability of a future result to be predicted by the model and is also used to verify the model accuracy.

The created model may fit perfectly to the training data set but may not qualify as a good model to represent the output for the given process at other points. For this reason, the verification data set is created so that the points are at different locations than the original sample. It is a good idea not to use the verification set for training, since it will defeat the purpose of testing the metamodel on totally unbiased points. If the verification RMSE and $R^2$ values do not differ very much from the training values, then the model has been trained correctly, otherwise it has not.

\subsection{Resolving Over-fitting in iVAMS 2.0}

An ML model (including ANN or deep neural network (DNN)) is over-fitted or inflated if the accuracy of the DNN model is better than the training dataset \cite{Bilbao_INTELCIS_2017}. In other words, the ANN/DNN architecture may be more complex than it is required for a specific problem. Different solutions such as the use of different datasets and reduction of the complexity of the ANN/DNN can be deployed. In ML modeling, over-fitting is the phenomenon where the ANN becomes worse instead of improving after a certain point during training when it is trained to as low errors as possible \cite{Garitselov_VLSID2012-PLL}. This is because excessive training or a large amount of neurons in the hidden layer may make the network memorize the training patterns and stop adjusting the weights. There are several methods to avoid over-fitting. One method is regularization which tries to limit the complexity of the network such that it is unable to learn peculiarities. Another method is early stopping which aims to stop training at the point of optimal generalization.

%%%%%%%%%%%%%%%%%%%%%%%%%%%%%%%%%%%%%%%%%%%%
\section{Comparative Perspective with Prior Research}
\label{sec_prior}
% discuss citations first
% then black-, white-, and grey-box

The current literature is rich in research attempting to speed up the design process of complex AMS circuits. Design space exploration approaches from high level descriptions of analog circuits are given in \cite{Doboli:2003:CADICS}. Posynomial modeling (a posynomial is a special type of polynomial \cite{Duffin1967a}) for gate sizing is presented in \cite{Roy:2007:CADICS}. A layout-aware modeling approach for analog synthesis is given in \cite{Pradhan:2008:LAS}. A single manual design iteration design flow is proposed in \cite{GhaiTVLSI2009Sep} for fast design optimization of VCOs. In \cite{KarabogaINISTA2011}, an ABC algorithm is used to determine the different parameters of a Schottky barrier diode.
In \cite{DelicanSM2ACD2010}, the ABC optimization algorithm is investigated considering the transient performance of a CMOS inverter circuit. Swarm intelligence algorithms including ABC and particle swarm optimization as well as differential evolution and its variant techniques are used for analog circuit optimization using a CMOS Miller operational transconductance amplifier in \cite{SabatNaBIC2009}.

The following are selected research works that have applied ANNs in VLSI design. In \cite{Wolfe2003}, the author shows that ANNs can be used for circuit analysis. In \cite{Linkun2009}, the authors introduce the creation of ANNs for estimating the output of operational amplifiers from a high level perspective which does not account for parasitics. In \cite{Zhongliang2004VLSID}, optimal and Hopfield binary ANNs are used for testing stuck-at-fault and delay faults in digital circuits. In \cite{Yazi2010EDAPS}, ANNs are trained on multi-dimensional mapping between geometrical variables and the values of independent circuit elements to predict the electromagnetic behavior of vias. In \cite{Zaabab1993}, the authors propose to speed up simulations by replacing repeated simulation data such as polynomial and look up models with well trained ANNs. In \cite{Macii1996CDT}, a Hopfield ANN model is used to represent digital circuit behavior. In \cite{Xu2003}, a feed-forward dynamic neural network model is developed for amplifier and mixer circuits directly from input-output large-signal measurements, without having to rely on the internal details of the circuit. In \cite{Zhang2009}, ANNs are used for electromagnetic susceptibility analysis and optimization of electronic devices.

The iVAMS proposed in this work is built upon neural network metamodels. These intelligent surrogate models facilitate fast block-level OP-AMP optimization and accurate parameterized behavioral modeling which enables efficient system-level design exploration. Modeling nano-CMOS AMS circuits using metamodeling techniques is becoming popular \cite{gar_2012}. Metamodeling OP-AMP performance metrics using neural networks was studied in \cite{wol_2003} for their capability of capturing the highly nonlinear nature of analog circuits. More recently, the need for efficient system-level design exploration kindled interest on parameterized macromodeling. In \cite{wang_2007}, an OP-AMP design space was partitioned and the sub-regions were approximated with local low-order polynomials to overcome the strong nonlinearity in large design spaces.

In this work, we apply ANN metamodels to OP-AMP parameterized macromodels (meta-macromodels) to avoid design space partitioning. Moreover, this ANN based meta-macromodel is implemented in Verilog-AMS for practical behavioral simulations. Using neural networks for behavioral modeling has been explored in the literature, e.g., \cite{liu_2002,jon_2011}. Nevertheless, actual implementation of such neural networks for AMS circuits in a hardware description language (HDL) has not been presented. From other related research, the closest one to this work is \cite{suz_2000}. However, it implemented a neural network with built-in training algorithm but not a behavioral model for any common analog circuit. Thus from our discussion, the current contribution of iVAMS is unique and practical.

%%%%%%%%%%%%%%%%%%%%%%%%%%%%%%%%%%%%%%%%%%%%%
\section{iVAMS 2.0 Case Study: An Operational Amplifier (OP-AMP)}
\label{sec_case_OP-AMP}

\begin{figure*}[htbp]
\centering
\includegraphics[width=0.85\textwidth]{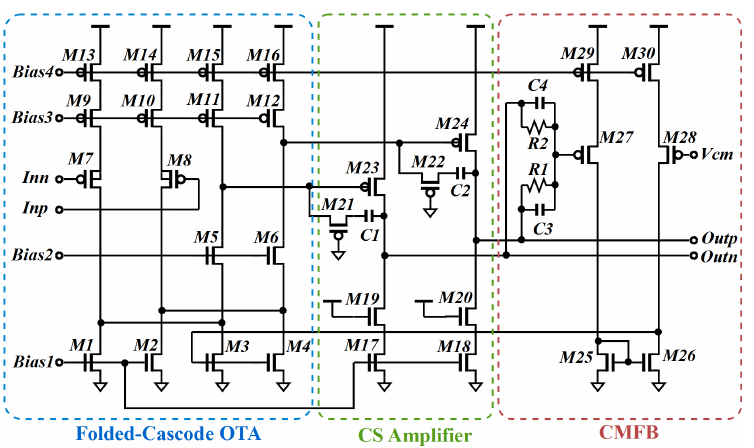}
\caption{The schematic of the OP-AMP case study \cite{Zheng_ASAP2013}.}
\label{fig_sch}
\end{figure*}

%An op amp is an important analog blocks in an AMS system. Thus
We apply iVAMS to the fully differential op amp shown in Fig.~\ref{fig_sch}, which is based on a 90nm CMOS process and a 1-V supply. It consists of a folded-cascode operational transconductance amplifier (OTA), a common-source (CS) amplifier, and a common-mode feedback (CMFB) circuit. Section~\ref{sec_case_gen} describes the generation of artificial neural network metamodels for the OP-AMP along with their accuracy evaluations. Section~\ref{sec_case_optim} presents a multi-objective OP-AMP optimization using the generated iVAMS PMMs. A meta-macromodel in Verilog-AMS for the OP-AMP is constructed in Section~\ref{sec_case_macro} using the iVAMS CPMs.

\subsection{OP-AMP iVAMS 2.0 Metamodel Generation}
\label{sec_case_gen}

The performance metrics of interest for this OP-AMP are the open-loop DC gain ($A_{0}$), bandwidth ($BW$), phase margin ($PM$), slew rate ($SR$), and power dissipation ($PD$). Metamodels for these metrics were generated for iVAMS block-level optimization. In order to construct a meta-macromodel for system-level design exploration, metamodels for the circuit parameters used in the macromodel are also required. These circuit parameters include the transconductance $g_{m}$, and the positive and negative maximum available currents $I_{p}$ and $I_{n}$ of the OP-AMP input stage. As a specific example, an ANN architecture with a single hidden layer is chosen for the metamodels \cite{cyb_1989}. 
%With sufficient neurons in the hidden layer, this architecture can approximate any function \cite{cyb_1989}.

%\begin{figure*}[htbp]
%\centering
%\includegraphics[width=0.70\textwidth]{nn_arch}
%\caption{Artificial Neural network (ANN) modeling.}
%\label{fig_nnarch}
%\end{figure*}

The input layer, $\mathbf{x}$, is a vector of the OP-AMP design variables which include the bias current and the transistor widths and lengths. There are thirty transistors in the OP-AMP design. By properly grouping the transistors, we have sixteen design variables ($N = 16$). For example, the current mirror devices $M1$--$M4$ in Fig.~\ref{fig_sch} should all have the same size, therefore two variables, instead of eight, are used. In this work, the hidden layer consists of four neurons ($M = 4$) with hyperbolic tangent function as the activation function $f_1$. The output layer is a single neuron employing a linear activation function $f_2$. The model output $\hat y$ is one of the performance metrics or circuit parameters. $W_1$ is a matrix composed of the weights of the connections from the design variables in the input layer to the neurons in the hidden layer. Similarly, $W_2$ is formed by the weights of the connections from the hidden layer to the output layer. Additional control on each neuron is through bias $b_{ij}$ ($i = 1, 2$ and $j = 1, 2, ..., M$). \textbf{The artificial neural networks were trained using 500 samples with Bayesian Regulation training}. It may be noted that, while different mathematical function can be be used for metamodeling, an ANN used in the current paper for it Verilog-AMS integration as the ANN is accurate and capable of modeling complex systems \cite{Garitselov_VLSID2012-PLL}. The number of simulations needed to generate metamodels is significantly lower than those needed for actual netlist simulation \cite{gar_2012}. At the same time, design exploration over metamodels is significantly faster than the simulations over netlists in SPICE. Thus, in summary the ANN metamodel integrated Verilog-AMS is ultra-fast while maintaining circuit-level accuracy.

%\subsection{OP-AMP iVAMS 2.0 Accuracy}
%\label{sec_case_accuracy}

In order to evaluate the accuracy of the iVAMS metamodels, \textbf{a verification set consisting of 2000 samples} of design variable and circuit response pairs were generated. It may be needed that this sampling is for verification purposes only and need not be used when iVAMS is used in design flow. The output $\hat y$ computed by the metamodel is compared with the ``true'' output $y$ obtained from the transistor-level simulation. To comparatively assess the proposed iVAMS ANN metamodels, we generated second-order polynomial (PO) metamodels and compared their accuracy using the coefficient of determination ($R^2$), relative maximum absolute error (RMAE), root relative square error (RRSE), and root mean square error (RMSE). RMSE measures the overall error in terms of the unit of the modeled output. RRSE expresses this overall error relative to the standard deviation of the true output \cite{yel_2012}. RMAE assesses the model locally and finds the maximum relative error \cite{jin_2001}. $R^2$ measures the model quality globally with respect to the variance of the true output. An accurate model should have small RMAE, RRSE, and RMSE. Its $R^2$ value should be close to 1. The results are listed in Table~\ref{tab_accuracy}.

\begin{table}[t]
\renewcommand{\arraystretch}{1.3}
% if using array.sty, it might be a good idea to tweak the value of
% \extrarowheight as needed to properly center the text within the cells
\caption{Accuracy of The OP-AMP iVAMS Metamodels \cite{Zheng_ASAP2013}.}
\label{tab_accuracy}
\centering
% Some packages, such as MDW tools, offer better commands for making tables
% than the plain LaTeX2e tabular which is used here.
\begin{tabular}{|c|c|c|c|c|c|}
\hline
\multicolumn{2}{|c|}{\bfseries Metamodel} & \multicolumn{4}{|c|}{\bfseries Accuracy Metric} \\
\hline
\bfseries Output & \bfseries Type & \bfseries $R^2$& \bfseries RMAE & \bfseries RRSE & \bfseries RMSE \\ \cline{2-5}
\hline \hline
\multirow{2}{*}{$\boldsymbol{A_0}$}
& NN & 0.959 & 1.324 & 0.202 & 41.93 V/V \\ \cline{2-6}
& PO & 0.973 & 1.044 & 0.163 & 33.78 V/V \\ \hline
\multirow{2}{*}{$\boldsymbol{BW}$}
& NN & 0.987 & 0.894 & 0.116 & 2.12 kHz  \\ \cline{2-6}
& PO & 0.986 & 0.965 & 0.117 & 2.14 kHz \\ \hline
\multirow{2}{*}{$\boldsymbol{PM}$}
& NN & 0.901 & 2.161 & 0.317 & 4.99$^o$  \\ \cline{2-6}
& PO & 0.348 & 4.466 & 0.807 & 12.70$^o$ \\ \hline
\multirow{2}{*}{$\boldsymbol{SR}$}
& NN & 0.989 & 0.483 & 0.105 & 0.292 mV/ns  \\ \cline{2-6}
& PO & 0.985 & 0.662 & 0.119 & 0.332 mV/ns  \\ \hline
\multirow{2}{*}{$\boldsymbol{P_D}$}
& NN & 0.996 & 0.523 & 0.062 & 8.306 $\mu$W \\ \cline{2-6}
& PO & 0.980 & 1.314 & 0.141 & 18.817 $\mu$W  \\ \hline
\multirow{2}{*}{$\boldsymbol{g_m}$}
& NN & 0.999 & 0.106 & 0.018 & 1.769 $\mu$A/V \\ \cline{2-6}
& PO & 0.999 & 0.101 & 0.021 & 1.973 $\mu$A/V \\ \hline
\multirow{2}{*}{$\boldsymbol{I_p}$}
& NN & 0.991 & 0.675 & 0.095 & 0.311 $\mu$A\\ \cline{2-6}
& PO & 0.729 & 3.407 & 0.521 & 2.506 $\mu$A  \\ \hline
\multirow{2}{*}{$\boldsymbol{I_n}$}
& NN & 0.994 & 0.494 & 0.080 & 0.261 $\mu$A \\ \cline{2-6}
& PO & 0.749 & 3.727 & 0.501 & 2.412 $\mu$A \\ \hline
\end{tabular}
\end{table}

The ANN metamodels achieve higher accuracy overall except for $A_0$. For some circuit responses (e.g., $PM$, $I_p$, and $I_n$), the PO metamodels deliver poor accuracy, while the ANN metamodels attain reasonable accuracy. Another advantage of the ANN metamodels over the PO metamodels is that their accuracy can be further improved by adding more neurons to the hidden layer, without drastically increasing the model complexity. All ANN metamodels in this work employ a 4-neuron hidden layer for simplicity. In practice, adaptively adjusting the hidden layer size is recommended to find the optimal model.

\subsection{OP-AMP iVAMS 2.0 Meta-Macromodel Construction}
\label{sec_case_macro}

\begin{figure}[htbp]
	\centerline{
		\includegraphics[width=0.80\textwidth]{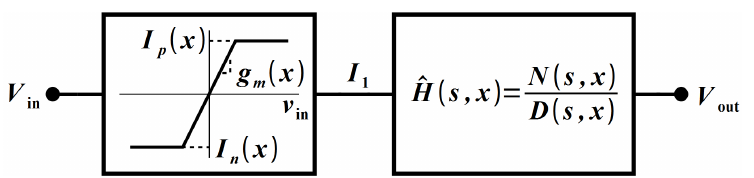}}
	\caption{The iVAMS 2.0 OP-AMP meta-macromodel \cite{Zheng_ASAP2013}.}
	\label{fig_symb}
\end{figure}

An OP-AMP meta-macromodel can facilitate fast system-level design exploration. Meta-macromodel construction starts with macromodel selection. Some OP-AMP macromodels that can be used include the structural model in \cite{gom_1995}, the linear time-invariant (LTI) model in \cite{wang_2007}, and the symbolic model in \cite{zha_2011}. We adopted a symbolic model similar to the one in \cite{zha_2011} since it not only models the OP-AMP small-signal behavior but also its large-signal behavior such as slew-rate limitation. This model, combined with the iVAMS CPMs to form the OP-AMP meta-macromodel, is shown in Fig.~\ref{fig_symb}.

The two-stage model in Fig.~\ref{fig_symb} takes into account the slew-rate effect due to the limited maximum available positive and negative currents $I_p$ and $I_n$. These circuit parameters, together with the transconductance of the first stage $g_m$ and the OP-AMP small-signal function $\hat H(s)$, are functions of the design variables. They are estimated using the iVAMS CPMs. These CPMs all have the same ANN architecture. With $f_1$ being a hyperbolic tangent function and $f_2$ being a pure linear function, this architecture can be expressed mathematically as:
\begin{equation}
\label{eq_nn}
\hat y = b_{21} + \sum_{j = 1}^{M} w_{2,j} \cdot \tanh\left(b_{1 j} + \sum_{i = 1}^{N} w_{1,ij} \cdot x_i \right),
\end{equation}
where $w_{1,ij} \in {W_1}$ and $w_{2,j} \in {W_2}$. Implementing this equation in Verilog-AMS is an essential step of building the iVAMS meta-macromodel. An example is shown in Algorithm~\ref{ALG:code}. After training the neural networks for the CPMs, the neural network weights and biases are stored in text files. In the \verb+initial+ block of the OP-AMP Verilog-AMS module, these weights and biases are read from the files, and the function \verb+nn_metamodel+ computes the circuit parameter values for the meta-macromodel. The computed circuit parameter values are used in an \verb+analog+ process in the Verilog-AMS module to realize the model in Fig.~\ref{fig_symb}. The OP-AMP small-signal function can be described using a function such as \verb+laplace_nd+.

\begin{algorithm}[htbp]
	\centering
	%\begin{figure}[htbp]
	%\rule{3.4in}{0.8pt}
	%\begin{minipage}[l]{3.4in}
	\caption{Verilog-AMS code for the OP-AMP artificial neural network (ANN) metamodels \cite{Zheng_ASAP2013}.}
	\label{ALG:code}
	%\begin{algorithmic}[1]
	%\footnotesize
	%\hspace{3.4in}
	\begin{verbatim}
	function real nn_metamodel;
	integer w1, w2, b1, b2, i, j, readfile,...;
	real w, b, v, u;
	// Read metamodel weights and bias from 
	// text files w1, w2, b1, and b2.
	begin
	w1 = $fopen("w1.txt", "r");
	w2 = $fopen("w2.txt", "r");
	b1 = $fopen("b1.txt", "r");
	b2 = $fopen("b2.txt", "r");
	v = 0.0;
	for (j = 0; j < nl; j = j + 1)
	begin
	u = 0.0;
	for (i = 0; i < size_x; i = i + 1)
	begin
	readfile = $fscanf(w1, "%e", w);
	u = u + w * x[i];
	end
	readfile = $fscanf(w2, "%e", w);
	readfile = $fscanf(b1, "%e", b);
	v = v + w * tanh(u + b);
	end
	readfile = $fscanf(b2, "%e", b);
	nn_metamodel = v + b;
	$fclose(w1);
	$fclose(w2);
	$fclose(b1);
	$fclose(b2);
	end
	endfunction
	\end{verbatim}
	%\end{minipage}
	%\rule{3.4in}{0.8pt}
	%\hspace{3.4in}
	%\caption{Verilog-AMS code for the op-amp neural network metamodels.}
	%\label{fig_code}
	%\end{figure}
	%\end{algorithmic}
\end{algorithm}

\begin{figure}[htbp]
	\centering
	\includegraphics[width=0.65\textwidth]{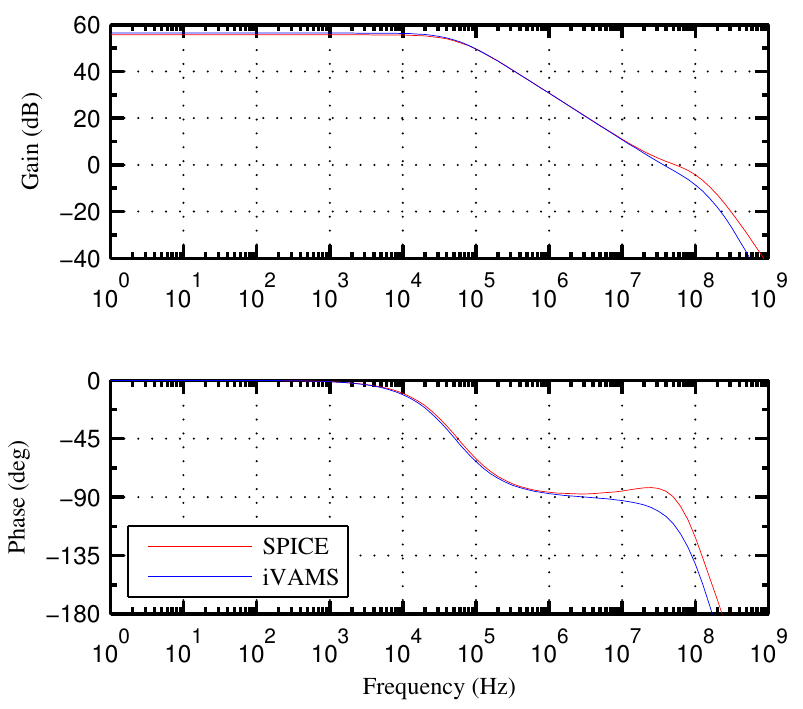}
	\caption{AC analyses of the OP-AMP \cite{Zheng_ASAP2013}.}
	\label{fig_acplot}
\end{figure}

\begin{figure}[htbp]
	\centering
	\includegraphics[width=0.65\textwidth]{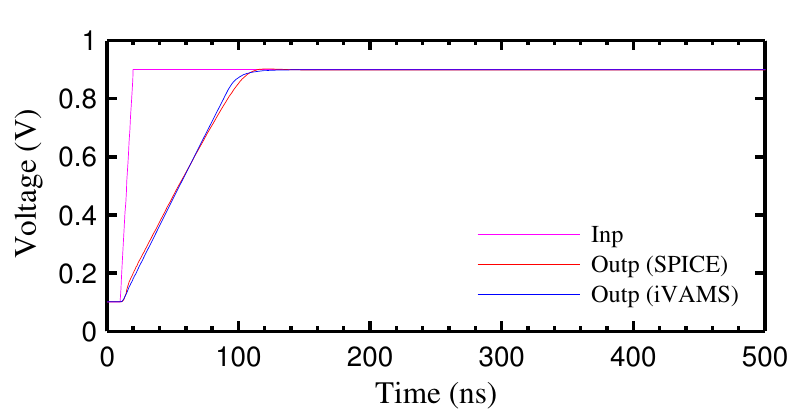}
	\caption{Transient simulations of the OP-AMP with a step input.}
	\label{fig_stepplot}
\end{figure}

\begin{figure}[htbp]
	\centering
	\includegraphics[width=0.65\textwidth]{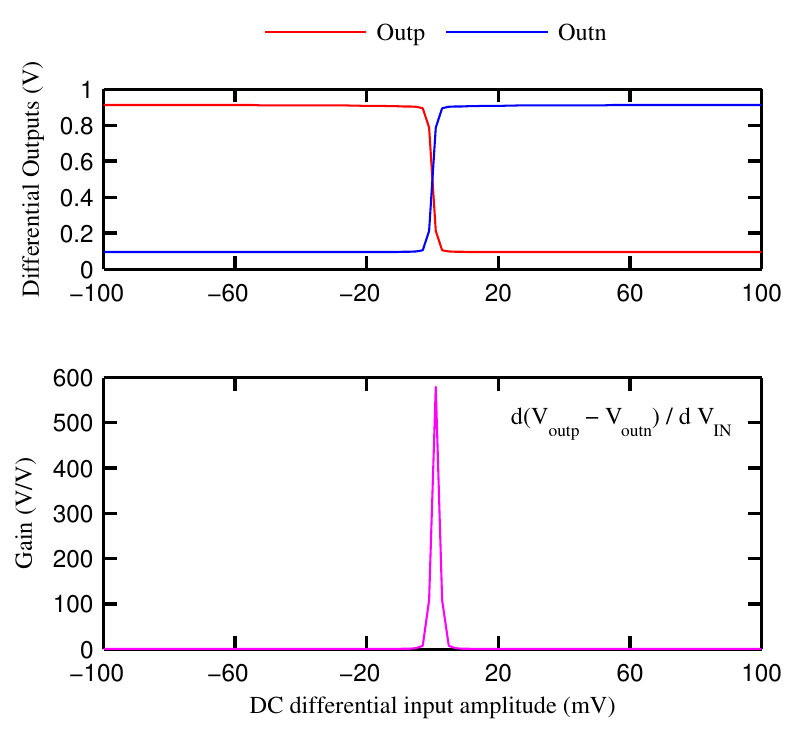}
	\caption{The DC behavior and gain of the OP-AMP.}
	\label{fig_dcplot}
\end{figure}

The quality of the meta-macromodel relies on the accuracy of the CPMs and the suitability of the selected macromodel. In order to validate the iVAMS meta-macromodel, simulations using the constructed Verilog-AMS module were compared with those using the SPICE model. The design variable values were set to be those of the selected optimal design in Section~\ref{sec_case_optim}. The results of the AC analysis are shown in Fig.~\ref{fig_acplot}. The difference seen in the frequency responses can be reduced by adaptively adjusting the neural network hidden layer sizes instead of using a fixed value. The OP-AMP step response is shown in Fig.~\ref{fig_stepplot}. The
DC behavior and gain are shown in Fig. \ref{fig_dcplot}. These simulations demonstrate that an effective OP-AMP meta-macromodel has been constructed.

\subsection{OP-AMP iVAMS 2.0 Multi-objective Optimization}
\label{sec_case_optim}

In this section, we demonstrate block-level optimization for the OP-AMP design using the iVAMS PMMs. Here, the OP-AMP is required to have high slew rate while consuming minimum power. Thus \textbf{the optimization problem is to maximize $\mathbf{SR}$ and to minimize $\mathbf{PD}$, constrained by the requirements for $\mathbf{A_0}$, $\mathbf{BW}$, and $\mathbf{PM}$}. An effective approach to this problem is to find the Pareto front (PF) that consists of a set of non-dominated solutions for the optimization problem. The designer can then select one design from this solution set to implement. Metaheuristic algorithms have shown promising results in solving multi-objective analog circuit optimization problems. For example, a particle swarm optimization (PSO) was used in \cite{fak_2010}; a memetic search based algorithm was used in \cite{liu_2010}; and a differential evolution (DE) based algorithm is used in \cite{liu_2011}. In this study, a \textbf{multi-objective firefly algorithm (MOFA)} belonging to the same category is used. It mimics the behavior of tropic firefly swarms that are attracted toward flies with higher flash intensity. Preliminary studies show that its performance could surpass existing established multi-objective algorithms \cite{yan_2012}. The iVAMS-assisted OP-AMP MOFA proposed in this work in shown in Algorithm~\ref{alg_mofa}.

\begin{algorithm*}[t]
\caption{iVAMS-Assisted Firefly based OP-AMP Multi-objective Optimization.}
\label{alg_mofa}
\begin{algorithmic}[1]
\STATE Current iteration $t \gets 0$;
\STATE Initialize a vector of $K$ designs $\mathbf{X} = \{\mathbf{x}_1, \mathbf{x}_2, ..., \mathbf{x}_K\}$;
\WHILE{$t < t_{max}$}
	\STATE Evaluate OP-AMP performance for all designs in $\mathbf{X}$ using iVAMS PMMs;
	\STATE Find non-dominated designs $\mathbf{X}_{nd} \subset \mathbf{X}$;
	\FOR{$i=1$ to $K$}
		\IF {No non-dominated designs are found}
			\STATE Generate random weight $w \in [0,1]$;
			\STATE Find a best design $\mathbf{x}_{*} \in \mathbf{X}$ that maximizes $\psi(\mathbf{x}) = (1 - w) \cdot \widehat{SR}(\mathbf{x}) - w \cdot \widehat{PD}(\mathbf{x}) $;
			\STATE Compute a move vector $\Delta \mathbf{x}_i$ for $\mathbf{x}_i$ toward $\mathbf{x}_*$;
		\ELSE
			\STATE Compute a move vector $\Delta \mathbf{x}_i$ for $\mathbf{x}_i$ toward $\mathbf{X}_{nd}$;
		\ENDIF
		\STATE Constraint$_1 \gets  A_0(\mathbf{x}_i + \Delta \mathbf{x}_i) > A_{0min}$;
		\STATE Constraint$_2 \gets  BW(\mathbf{x}_i + \Delta \mathbf{x}_i) > BW_{min}$;
		\STATE Constraint$_3 \gets  PM(\mathbf{x}_i + \Delta \mathbf{x}_i) > PM_{min}$;
		\IF {Not all constraints are satisfied}
			\STATE Generate a new move;
		\ENDIF
	\ENDFOR
	\STATE Let $\Delta \mathbf{X} = \{\Delta\mathbf{x}_1, \Delta\mathbf{x}_2, ..., \Delta\mathbf{x}_K\}$;
	\STATE $\mathbf{X} \gets \mathbf{X} + \Delta\mathbf{X}$;
	\STATE $t \gets t + 1$;
\ENDWHILE
\end{algorithmic}
%\hspace{3.2in}
%}
%\end{minipage}
%\rule{3.4in}{0.8pt}
%\hspace{3.4in}
\end{algorithm*}

The goal of the algorithm is to find $K$ Pareto points that constitute the PF through a predetermined number of iterations, $t_{max}$. The algorithm starts with $K$ randomly generated designs. In each iteration, the performance of the $K$ OP-AMP designs are estimated using the iVAMS PMMs. If non-dominated designs are found, move vectors from each current design toward the non-dominated designs will be computed based on the attractiveness of these designs and the characteristic distance \cite{yan_2012}. If no non-dominated designs are found, a move vector toward a current best design determined by the combined weighted sum of the objectives, $\psi(\mathbf{x})$, is computed for each current design. $\widehat{SR}(\mathbf{x})$ and $\widehat{PD}(\mathbf{x})$ are the slew rate and power dissipation for a design $\mathbf{x}$ estimated using the iVMAS PMMs, and are normalized with respect to their own sample mean and standard variation. The current designs will be moved to new locations within the design space using the computed move vectors. The PMMs are then used to check whether the new designs satisfy the constraints. If not, new moves will be generated. The optimization specifications and an arbitrarily selected optimal design from the PF of a MOFA optimization are shown in Table~\ref{tab_mofa}.

\begin{table}[t]
\renewcommand{\arraystretch}{1.3}
% if using array.sty, it might be a good idea to tweak the value of
% \extrarowheight as needed to properly center the text within the cells
\caption{OP-AMP Design Optimization \cite{Zheng_ASAP2013}.}
\label{tab_mofa}
\centering
% Some packages, such as MDW tools, offer better commands for making tables
% than the plain LaTeX2e tabular which is used here.
\begin{tabular}{|c|c|c|c|c|}
\hline
& & \multicolumn{2}{c|}{\bfseries Selected Optimal Design} \\
\cline{3-4}
\bfseries  Performance & \bfseries  Constraint & \bfseries Predicted & \bfseries True \\
\hline \hline
$\boldsymbol{A_{0}}$ (dB) & $>$ 43 & {56.4} & {55.7} \\
{$\boldsymbol{BW}$} (kHz) & $>$ 50 & {56.8} & {56.7} \\
{$\boldsymbol{PM}$} (degree) & $>$ 70 & {81.9} & {88.5} \\
\hline
{} & \bfseries Objective & {} & {} \\
\hline
{$\boldsymbol{SR}$} (mV/ns) & Maximized & {5.54} & {5.49} \\
$\boldsymbol{P_{D}}$ (${\mu}$W) & Minimized & {85.11} & {85.77} \\ \hline
\end{tabular}
\end{table}

\begin{figure}[htbp]
\centering
\includegraphics[width=0.48\textwidth]{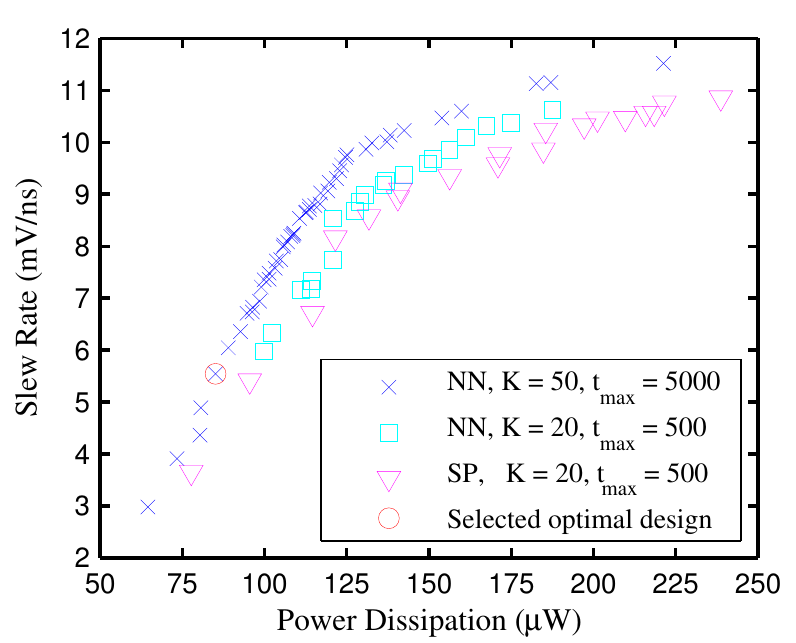}
\caption{Pareto fronts from MOFA OP-AMP optimization \cite{Zheng_ASAP2013}.}
\label{fig_mofa}
\end{figure}

The MOFA has been implemented in MATLAB. Three optimization runs have been performed. The first run is aimed at finding the true PF. In order to approximate the true PF, $K$ and $t_{max}$ have to be sufficiently large. In this run, $K = 50$ and $t_{max} = 5000$ were set experimentally. Although these numbers are large, the runtime for this optimization is small thanks to the neural network iVAMS PMMs. To compare the iVAMS-assisted MOFA (iVAMS-MOFA) with the same algorithm using SPICE models (SPICE-MOFA) to evaluate the OP-AMP performance, we may run a similar optimization. However, running MOFA using SPICE (SP) models with such large $K$ and $t_{max}$ would be extremely time consuming. Therefore, we alternatively decreased there numbers to $K = 20$ and $t_{max} = 500$ and performed two optimization runs to compare the speed of iVAMS-MOFA and SPICE-MOFA. These three optimizations are summarized in Table~\ref{tab_compare}. It can be seen that \textbf{the iVAMS-MOFA is 5580 times faster than the SPICE-MOFA}. The PFs generated by the three optimizations and the arbitrarily selected OP-AMP design are shown in Fig.~\ref{fig_mofa}.

\begin{table}[htbp]
\renewcommand{\arraystretch}{1.3}
% if using array.sty, it might be a good idea to tweak the value of
% \extrarowheight as needed to properly center the text within the cells
\caption{Comparison of MOFA OP-AMP Optimizations}
\label{tab_compare}
\centering
\begin{tabular}{|c|c|c|c|}
\hline
\bfseries Optimization \#
& \bfseries 1 & \bfseries 2 & \bfseries 3 \\ \hline \hline
\bfseries Model Type & ANN & ANN & SPICE \\
\bfseries Number of Pareto Points, $K$ & 50 & 20 & 20 \\
\bfseries Number of Iterations, $t_{max}$ & 5000 & 500 & 500 \\ \hline \hline
%\bfseries Runtime & 0.572 h & {84.635 s} & {131.185 h} \\
\bfseries Runtime & 0.57 h & {84.63 s} & {131.18 h} \\
\bfseries Normalized Speed & -- & {$\times$5580} & {1} \\
\hline
\end{tabular}
\end{table}

%\pagebreak
%%%%%%%%%%%%%%%%%%%%%%%%%%%%%%%%%%%%%%%%%%%%%
\section{iVAMS 2.0 Case Study: A Phase-Locked Loop (PLL)}
\label{sec_case_PLL}

This section presents ML modeling and optimization of a PLL through an ANN-based metamodel \cite{Garitselov_VLSID2012-PLL}. For non-polynomial metamodeling different architectures of neural networks are considered to perform trade-off analysis between speed and accuracy. As a practical demonstration of the use
of the non-polynomial neural network metamodels, the physical design optimization
of a 180 nm CMOS PLL is undertaken. A biologically inspired tool, the ABC algorithm is used for optimization of the PLL physical design that uses the metamodels instead of the actual circuit (i.e. the parasitic aware netlist).
It is demonstrated that the non-polynomial neural network metamodel assisted optimization is faster and more accurate compared to the polynomial metamodel.

\begin{figure}[htbp]
	\centering
	\includegraphics[width=0.68\textwidth]{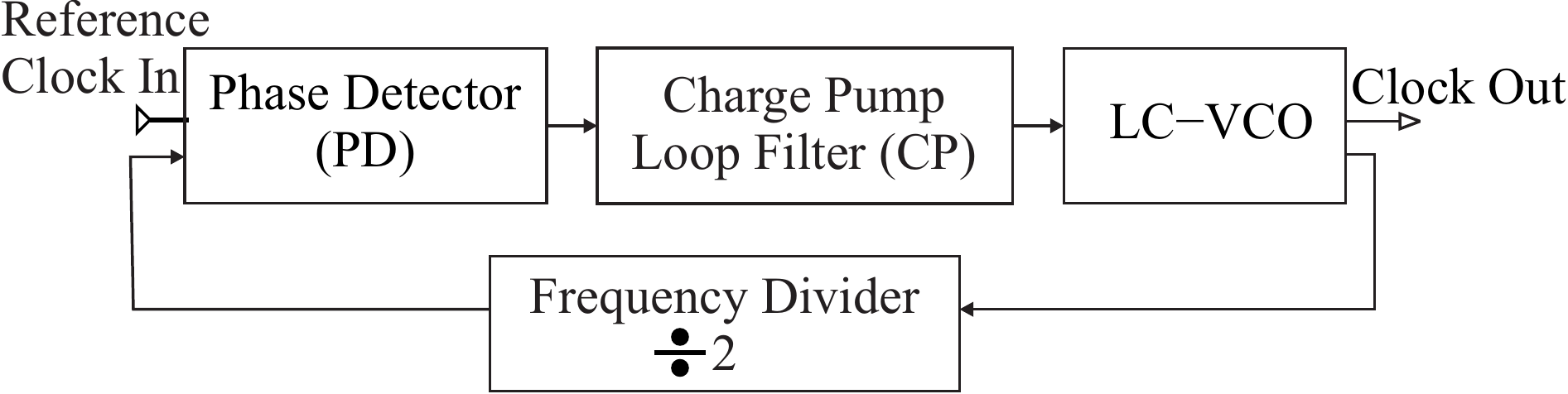}
	\caption{Block diagram of a phase locked loop (PLL) \cite{Garitselov_VLSID2012-PLL}.}
	\label{fig:PLL_Block_Diagram}
\end{figure}

\begin{figure}[htbp]
	\centering
	\includegraphics[width=0.35\textwidth]{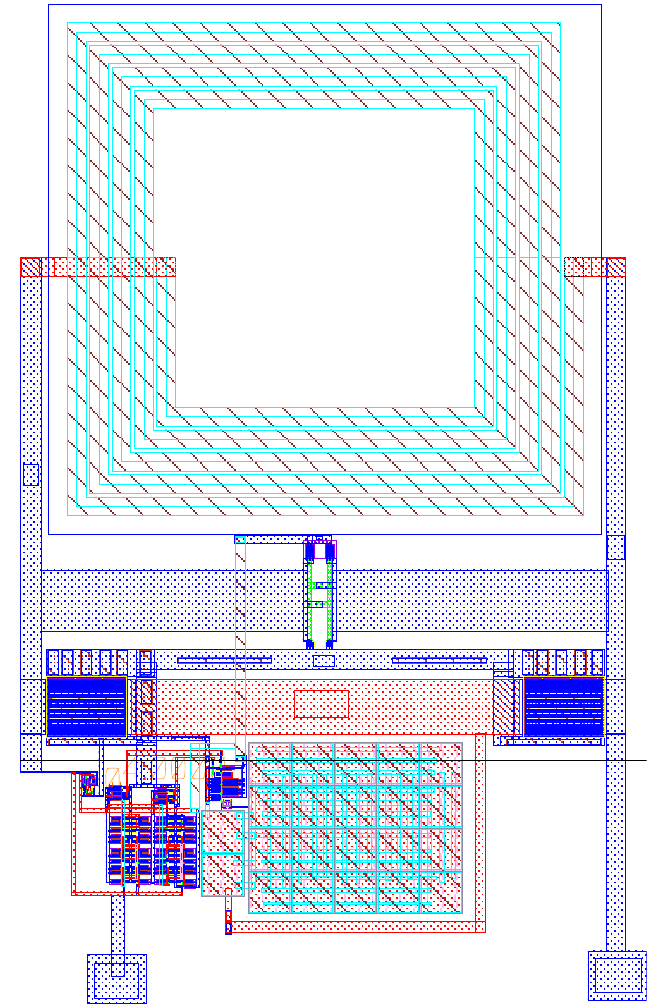}
	\centering
	\caption{Physical design of the PLL with area 525$\times$326 $\mu$m \cite{Garitselov_VLSID2012-PLL}.}
	\label{fig:layout}
\end{figure}

\begin{figure}[t]
	\centering
	\includegraphics[width=0.70\textwidth]{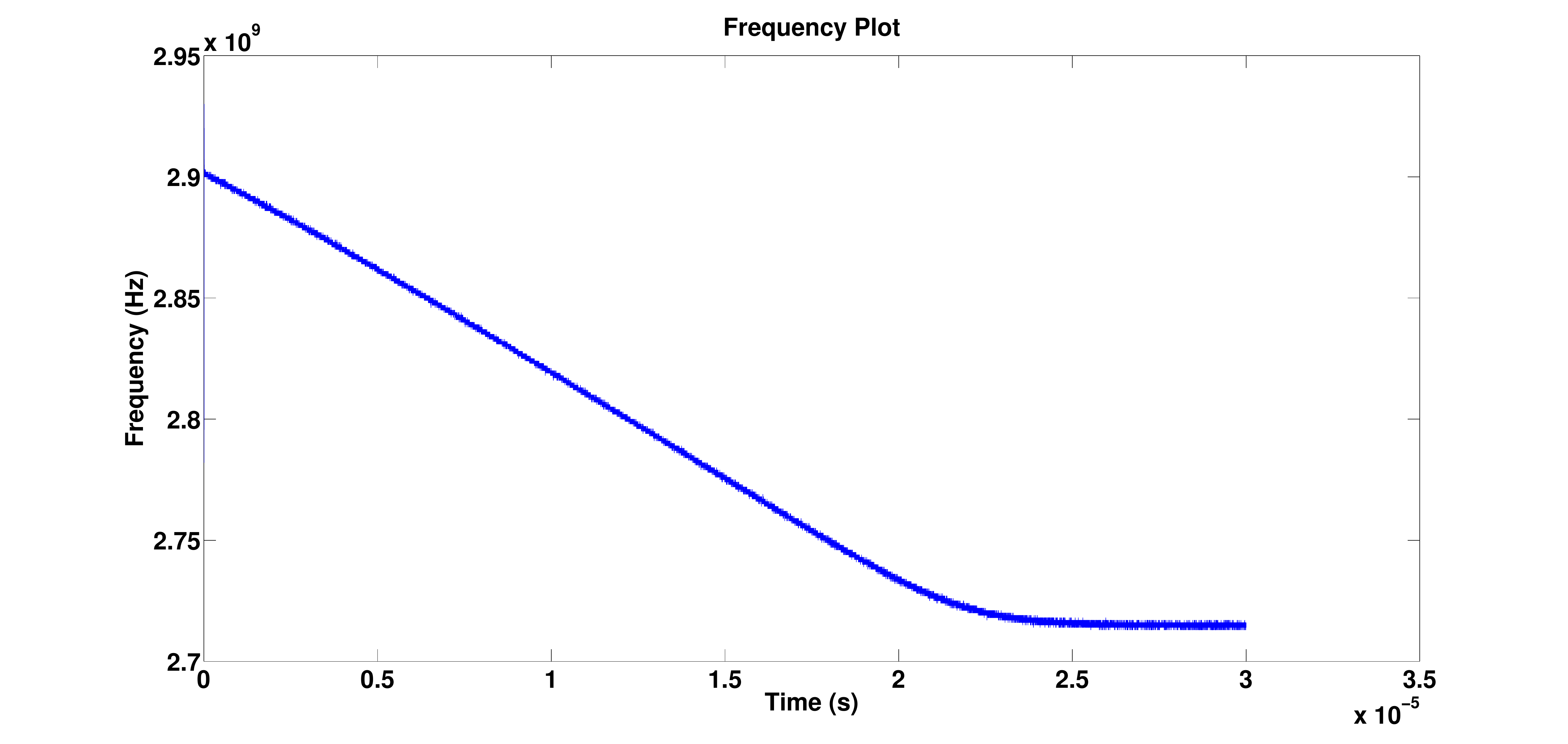}
	\centering
	\caption{Locking behavior of the PLL circuit \cite{Garitselov_VLSID2012-PLL}.}
	\label{fig:PLL_freq_plot}
\end{figure}

\begin{figure}[htbp]
	\centering
	\includegraphics[width=0.65\textwidth]{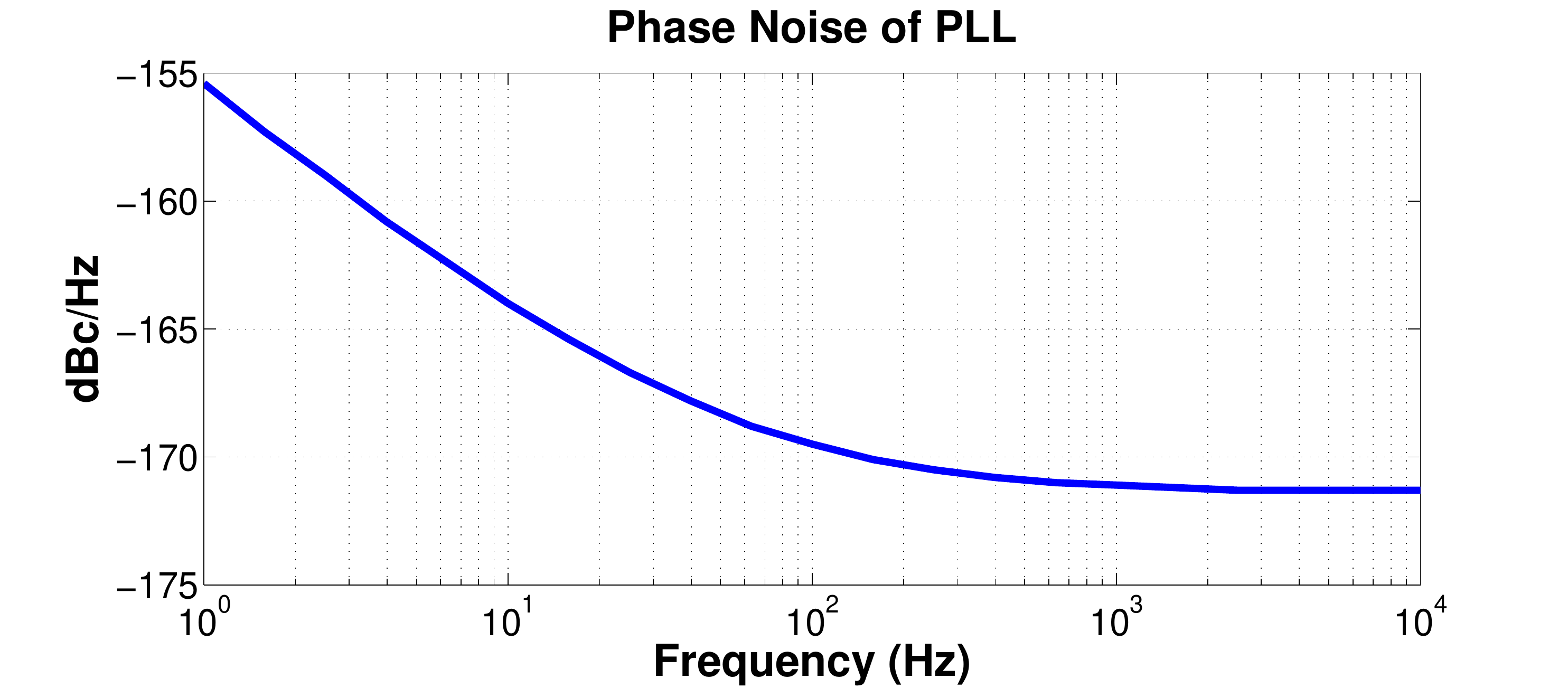}
	\centering
	\caption{Phase noise of the baseline PLL circuit \cite{Garitselov_VLSID2012-PLL}.}
	\label{fig:phasenoise}
\end{figure}

The phase locked loop (PLL) is a closed feedback loop system which is implemented as shown in Fig. \ref{fig:PLL_Block_Diagram}. The detailed baseline design of this circuit is discussed in \cite{Mohanty_Book_2015_Mixed-Signal,Garitselov_JOLPE2012Jun}. The physical design is shown in Fig. \ref{fig:layout}. A parasitic-aware netlist, including resistance (R), capacitance (C) and inductance (both self and mutual) (LK) is extracted from the layout. The netlist is then parameterized and used for simulations for the input data sets for each metamodel. Once the data are received from SPICE simulations,  they are processed by an external tool (Matlab). SPICE simulation results of the circuit are shown in Fig. \ref{fig:PLL_freq_plot}, which shows the frequency over time plot, with the PLL locking in 24.58 $\mu$s. The baseline phase noise diagram in Fig. \ref{fig:phasenoise} shows that the circuit has acceptable phase noise, (-163 dBc/Hz at 10 Hz offset). The average power consumption of the PLL is 9.28 mW.

\subsection{PLL iVAMS 2.0 Metamodel Generation}

Given that each SPICE simulation for the PLL circuit takes approximately 10 minutes to converge, the amount of simulation runs are limited. In this work 100 simulations for training and 30 simulations for verification have been chosen. Different architectures of ANNs are evaluated. For feed-forward networks two differentiable transfer functions (tanh - tansig, and logarithmic - logsig) are used for the hidden layer. In addition, the experimental results also consider the difference between raw regular input data in comparison to normalized and standardized input sets.

The verification data set is also chosen using LHS, but it is ensured that none of the points match the training set. After the neural network training is completed, the input values for the verification set are fed into the network and the RMSE value is calculated for the verification set. The $R^2$ values are calculated for training and verification sets for each combination of the above neural networks. Selected results are summarized in Tables \ref{tbl:freq_comp} and \ref{tbl:lockingtime_comp}. The statistics of the best created polynomial functions that were created from \cite{Garitselov_JOLPE2012Jun} are listed in the last rows of the tables for comparison purposes.

For comparison purposes, the data was fitted into partial polynomial equations. Since the full polynomial function would result in a very large number of coefficients for 21 variables, partial polynomial functions of order 1 through 6 are considered. Further, the stepwise regression method is used to filter out the coefficients that do not contribute to the function's outcome.

\begin{table*}[t]
	\caption{Frequency Non-Polynomial Metamodel Comparison of the PLL \cite{Garitselov_VLSID2012-PLL}.}
		\label{tbl:freq_comp}
\centering
		\begin{tabular}{|c|c|c|c|c|c|}
			\hline
			Function&Data Filtering & R$^2$-test & R$^2$- verification& RMSE & Neurons \\ \hline\hline
			logsig$\rightarrow$purelin & none & 0.802& 0.723&52.74 MHz &4 \\ \hline
			tansig$\rightarrow$purelin & none & 0.839& 0.713&51.24 MHz&3 \\ \hline
			radial$\rightarrow$purelin & none & 0.020& 0.490&81.51 MHz& \\ \hline
			\hline
			logsig$\rightarrow$purelin & minmax & 0.917 & 0.664& 48.89 MHz & 9 \\ \hline
			tansig$\rightarrow$purelin & minmax & 0.855 & 0.699& 53.65 MHz& 1 \\ \hline
			radial$\rightarrow$purelin & minmax & 0.844 & 0.712& 50.88 MHz& \\ \hline
			\hline
			logsig$\rightarrow$purelin & meanstd & 0.843 & 0.733 & 53.60 MHz& 1 \\ \hline
			tansig$\rightarrow$purelin & meanstd & 0.793 & 0.762 & 51.64 MHz& 5 \\ \hline
			radial$\rightarrow$purelin & meanstd & 0.848 & 0.749 & 48.97 MHz& \\ \hline
			\hline
			&Data Filtering&R$^2$&Order&RMSE&Number of Coefficients \\ \hline\hline
			polynomial&none&0.930&4&77.96 MHz&48 \\ \hline
		\end{tabular}
	
\end{table*}

\begin{table*}[htbp]
	\caption{Locking Time Non-Polynomial Metamodel Comparison of the PLL \cite{Garitselov_VLSID2012-PLL}.}
	\label{tbl:lockingtime_comp}
	\centering
		\begin{tabular}{|c|c|c|c|c|c|}
			\hline
			Function & Data Filtering & R$^2$-Test & R$^2$-Verification & RMSE & Neurons \\
			\hline\hline
			
			logsig$\rightarrow$purelin &none &0.828& 0.873 &1.30 $\mu$s &1\\ \hline
			tansig$\rightarrow$purelin &none &0.850& 0.723 &1.44 $\mu$s &9\\ \hline
			radial$\rightarrow$purelin &none &0.078& 0.830 &2.26 $\mu$s & \\ \hline
			\hline
			
			logsig$\rightarrow$purelin &minmax &0.826&0.870&1.29 $\mu$s&1\\ \hline
			tansig$\rightarrow$purelin &minmax &0.839&0.942&1.12 $\mu$s&10\\ \hline
			radial$\rightarrow$purelin &minmax &0.931&0.508&1.65 $\mu$s& \\ \hline
			\hline
			logsig$\rightarrow$purelin &meanstd &0.826&0.906&1.22 $\mu$s&2\\ \hline
			tansig$\rightarrow$purelin &meanstd &0.737&0.939&1.12 $\mu$s&3\\ \hline
			radial$\rightarrow$purelin &meanstd &0.963&0.691&1.23 $\mu$s& \\ \hline
			\hline
			&Data Filtering&R$^2$&Order&RMSE&Number of Coefficients \\ \hline\hline
			polynomial&none&0.877&4&1.91 $\mu$s&56 \\ \hline
		\end{tabular}
\end{table*}

From the data it is observed that neural networks with no standardization of the input data perform the worst. Even though polynomials show best results without standardization or normalization,  this is not the case for neural networks. Also, it can be concluded that ANNs perform better fitting for this circuit, mostly because of the non-linear flexibility of the ANNs. The data also demonstrates which architecture and normalization or standardization should be used, i.e. which has the best performance.

%\subsection{PLL iVAMS 2.0 Accuracy}

\subsection{PLL iVAMS 2.0 based Multi-objective Optimization}

\begin{figure}[htbp]
	\centering
	\includegraphics[width=0.70\textwidth]{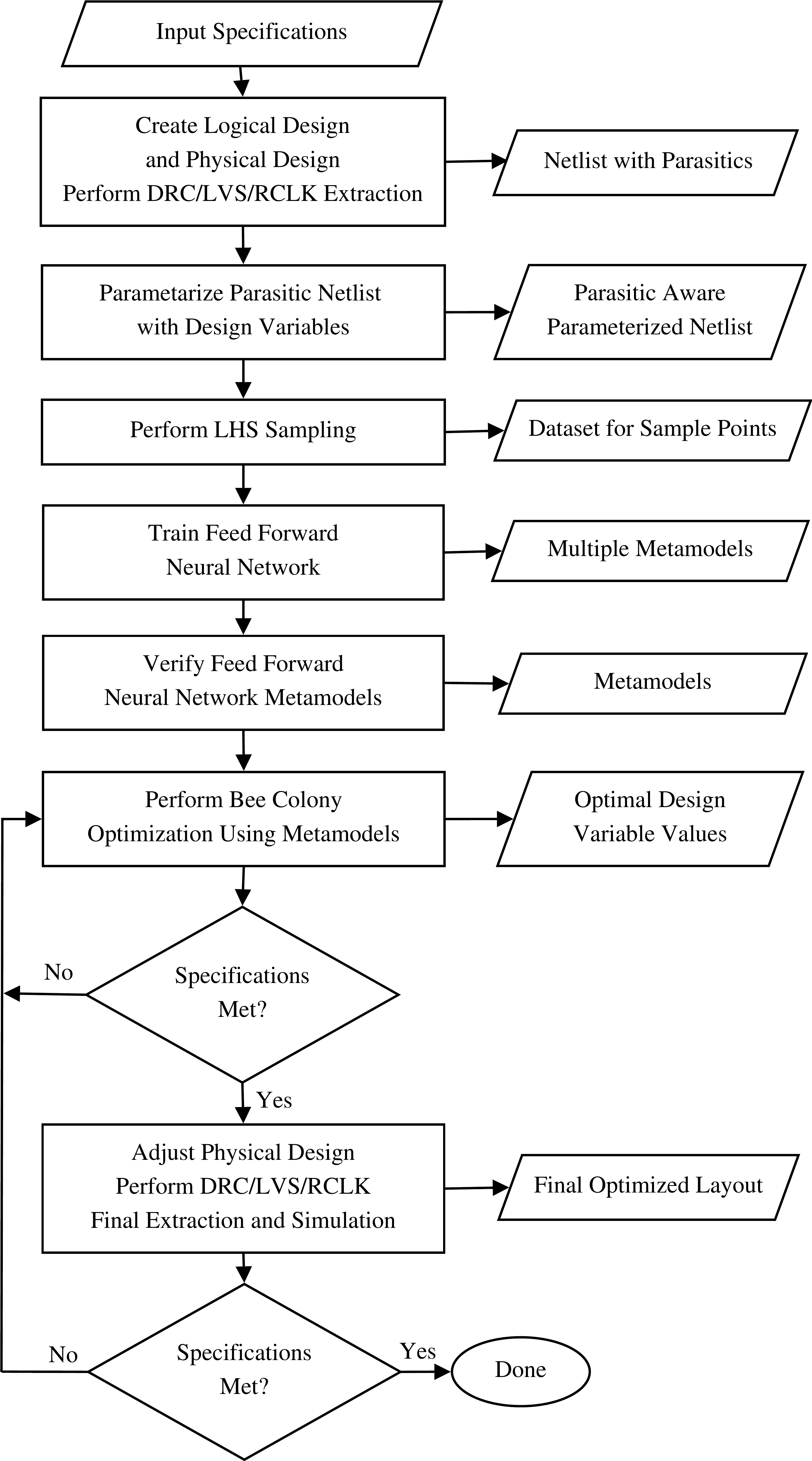}
	\caption{Proposed design flow utilizing feed-forward ANN metamodels and the ABC optimization algorithm.}
	\label{fig:PLL_Design_flow_ABC}
\end{figure}

The proposed design flow using non-polynomial metamodels is shown in Fig. \ref{fig:PLL_Design_flow_ABC}. Once the initial physical design netlist is parameterized, the design space is sampled using LHS. The ANN is trained from the sampled data. After the ANN is created and verified for accuracy, the optimization algorithm is applied to find the optimal parameters.
The non-polynomial metamodels are created for each output set of the design. Computationally expensive optimization algorithms can be applied using the fast non-polynomial metamodels as they are ultra fast compared to the actual netlist simulations. The optimized values are then used to adjust the initial physical layout to create the near optimal design. This design flow only uses two iterations for the physical design, at the beginning and the end. Overall, the design flow is as accurate as the parasitic-aware netlist of the circuit but ultra-fast due to the metamodel abstraction, which in turn minimizes the amount of time the designer needs to spend on the design optimization of the circuit.

The best (may be the most accurate or the fastest, depending on the requirement) non-polynomial metamodels from the previous section need to be selected for optimization. The optimization algorithm that is being used is the Artificial Bee Colony algorithm (ABC). ABC is the artificial representation of a bee colony behavior as bees try to find the best food source \cite{Karabora2009AMC,Hedayatzadeh2010ICACTE}. More information about the algorithm in the context of AMS circuit optimization can be found in our previous research \cite{Garitselov_JOLPE2012Jun}. Many metaheuristic algorithms are based on the swarm intelligence that can be found in nature. It was observed that in nature, honey bees are very efficient in finding enriched food sources by splitting tasks and communicating information amongst themselves. The artificial bee colony algorithm closely represents their behavior. All the bees in a colony can perform three different tasks: worker, onlooker and scout. The bees work independently and only communicate the information when they come back to the nest. When the bee is in the scout state, it travels to a random location and brings back information of how valuable the food source is. Bees in the worker state travel to the best known location and their whereabouts to collect food. The bees in the onlooker state, are located at the nest and based on the information of scout and worker bees decide to go and forage food or stay back.

This algorithm was found to be effective for use on AMS circuits with metamodels. As a specific objective and constraint optimization, the PLL circuit is characterized for output frequency, power, and locking time. A separate metamodel is created for each Figure-of-Merit (FoM) from the same sample set. Each single simulation calculates all FoMs so the number of simulations that are needed does not depend on the number of the metamodels that need to be created.

\begin{figure}[htbp]
	\centering
	\includegraphics[width=0.80\textwidth]{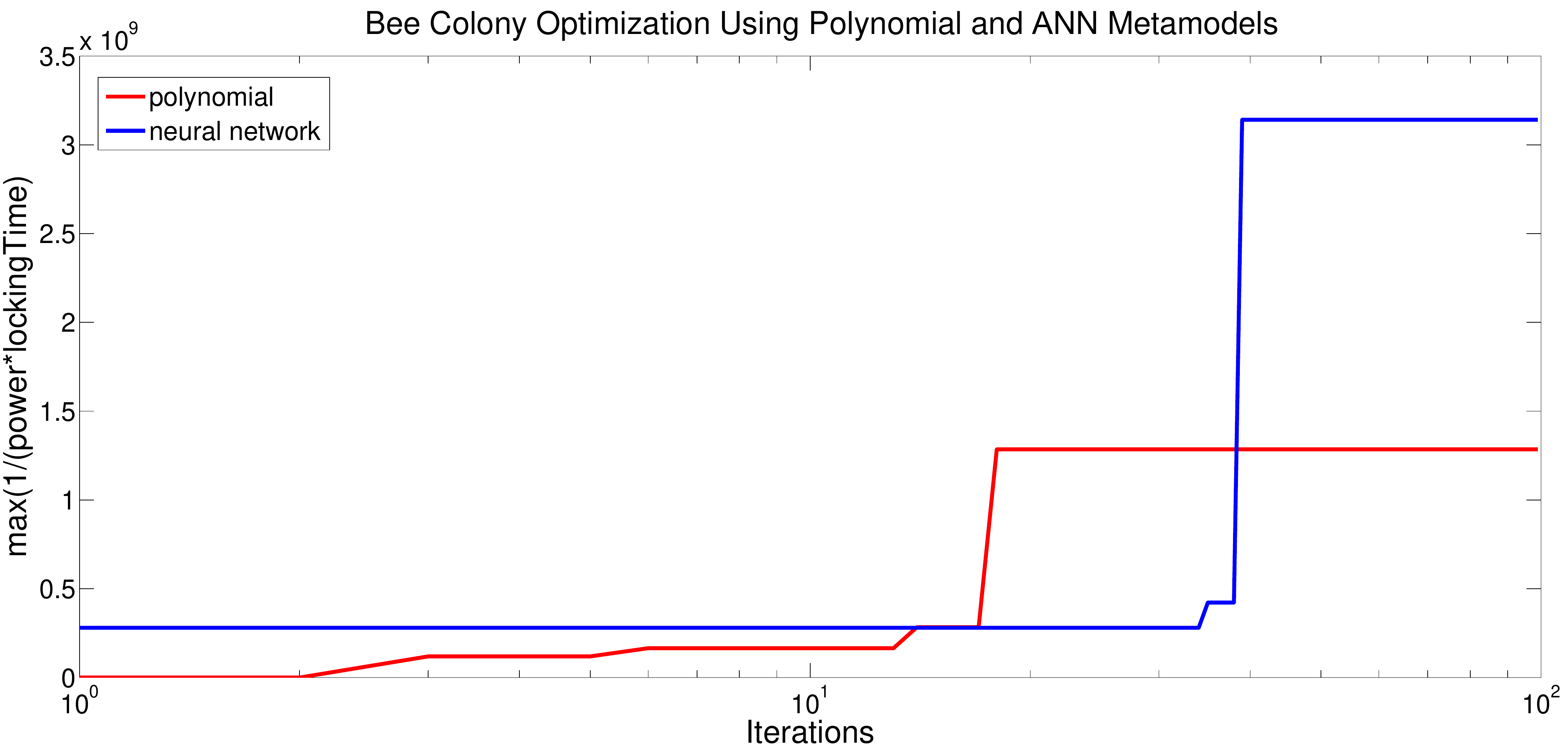}
	\caption{Artificial Bee Colony optimization conducted on polynomial and neural network metamodels for optimizing power and locking time within 0.5\% of 2.7 GHz frequency constraint \cite{Garitselov_VLSID2012-PLL}.}
	\label{fig:optimization}
\end{figure}

Figure \ref{fig:optimization} shows the progression of the FoM as the ABC optimization progresses. The frequency metamodel is used to filter (constrain) the results within 0.05\% of the required 2.7 GHz operational locking frequency of the PLL, which can be used in Multichannel Multipoint Distribution Systems (MMDS). The constraint narrows down the search criteria for the algorithm. The results of the neural-network metamodeling are compared with the polynomial metamodeling based approach. Table \ref{tbl:final_optim} shows the optimized values for both polynomial and ANN metamodels.

\begin{table}[htbp]
	\caption{PLL Circuit Characteristics after Optimization \cite{Garitselov_VLSID2012-PLL}.}
	\label{tbl:final_optim}
\centering
		\begin{tabular}{|c|c|c|}
			\hline
			PLL Characteristics        & Polynomial Metamodeling & ANN Metamodeling \\
			%&  &  \\ 
			 \hline \hline
			Power       &3.9 mW &3.9 mW \\ \hline
			Locking Time &8.476 $\mu$s&3.3147 $\mu$s \\ \hline
			Frequency   &2.6909 GHz&2.7026 GHz \\ \hline
		\end{tabular}
\end{table}

ANNs are reusable and can be used as a system of equations to accurately represent the needed output and are integrated in Verilog-AMS (to make iVAMS as shown in the previous section). In the case of complex AMS circuits like PLL, ANNs show on average 56\% increase in accuracy of prediction over the polynomial metamodels that have been generated from the same input data samples. In addition, neural network prediction, which is on average within 3.2\% of SPICE output, is enormously faster than SPICE simulation and is shown to find better solution during the optimization phase of design. Even though the circuit that this paper uses as an example is parameterized with 21 parameters, in future work higher and more complex circuits that can have hundreds of parameters will be investigated.

%%%%%%%%%%%%%%%%%%%%%%%%%%%%%%%%%%%%%%%%%%%%%%%%%%%%%%%%%
\section{Conclusion and Future Research}
\label{sec_conc}

The circuit-level accurate behavioral modeling framework called iVAMS has been presented in this article. The machine learning based artificial neural network integrated intelligent Verilog-AMS is called iVAMS 2.0 in this article (the predecessor iVAMS 1.0 is polynomial metamodel integrated Intelligent Verilog-AMS). The creation of an iVAMS module for an OP-AMP and PLL have  been discussed. The use of the iVAMS for block-level optimization has been demonstrated using a novel multi-objective firefly algorithm. Construction of parameterized behavioral models using iVAMS is also exemplified through an OP-AM case study. Future research includes enhancing iVAMS 2.0 with yield-estimation capability to address variability of nanometer circuits. For nanoelectronics design, manufacturing process variation is a key design issue as it affects device parameters and eventually chip yield.  This paper also presented the generation and usage of ANN metamodels of a PLL. The arificial bee colony algorithm with both non-polynomial and polynomial metamodels has been used for optimization. 

It is a fact that while ML models are attractive options for a variety of modeling applications, the training time, energy, and computational resource requirements are huge \cite{Mohanty_iSES_2018_Keynote,Neshatpour_FCCM_2018}. Thus, speeding up the training process as well as generating sufficiently complex models that represent the circuits and system behavior well while execute with minimal resources can be crucial. We intend to present iVAMS 3.0 that will incorporate a hierarchical modeling approach to speed up ML modeling.  The future research of iVAMS will include metamodeling while accounting for process variations and will model statistical parameters of the circuits and system characteristics.

In big application domains such as smart cities \cite{Mohanty_CEM_2016-Jul}, the large amount of data (aka bigdata) can be live (e.g. real-time sensors) as well as residing in various locations. However, the silicon data involved in the hardware design phase while complex, is not necessarily live or distributed. A designer typically deals with silicon data during or after the end of SPICE simulations, and of one design. Thus, distributed learning (including federated learning from Google) can have usage  for smart city kind applications \cite{McMahan_2017_Google-FL}. However, in the case of silicon bigdata involved in hardware design exploration hierarchical learning may be more useful to reduce the training time, which translates to reduction in design time and non-recurrent cost. However, the distributed paradigm for training of silicon data can also be explored to reduce training time and design cost.

%%%%%%%%%%%%%%%%%%%%%%%%%%%%%%%%%%%%%%%%%%%%%%%%%%%%%%%%%
\section*{Acknowledgments}

A preliminary version of this research appeared in the following peer-review conference \cite{Zheng_ASAP2013,Garitselov_VLSID2012-PLL}. It was also presented in work-in-progress poster session in DAC 2013 \cite{Zheng_DAC2013}. The authors would like to thank UNT graduates Dr. Geng Zheng and Dr. Oleg Garitselov for their help on earlier versions of this work.

%%%%%%%%%%%%%%%%%%%%%%%%%%%%%%%%%%%%%%%%%%%%%
%\bibliographystyle{bib/IEEEtran}
%\bibliographystyle{abbrv}
\bibliographystyle{bib/IEEEtran}
%\bibliographystyle{plain}
%\bibliographystyle{alpha}
%\bibliography{bib/IEEEabrv,bib/Bibliography_iVAMS2}


\vspace{-0.2cm}
%\pagebreak
\section*{Authors' Biographies}
\vspace{-0.2cm}

%\begin{description}
%	\item
\begin{wrapfigure}{l}{1.0in}
	\vspace{-0.5cm}
	\includegraphics[width=1.0in,keepaspectratio]{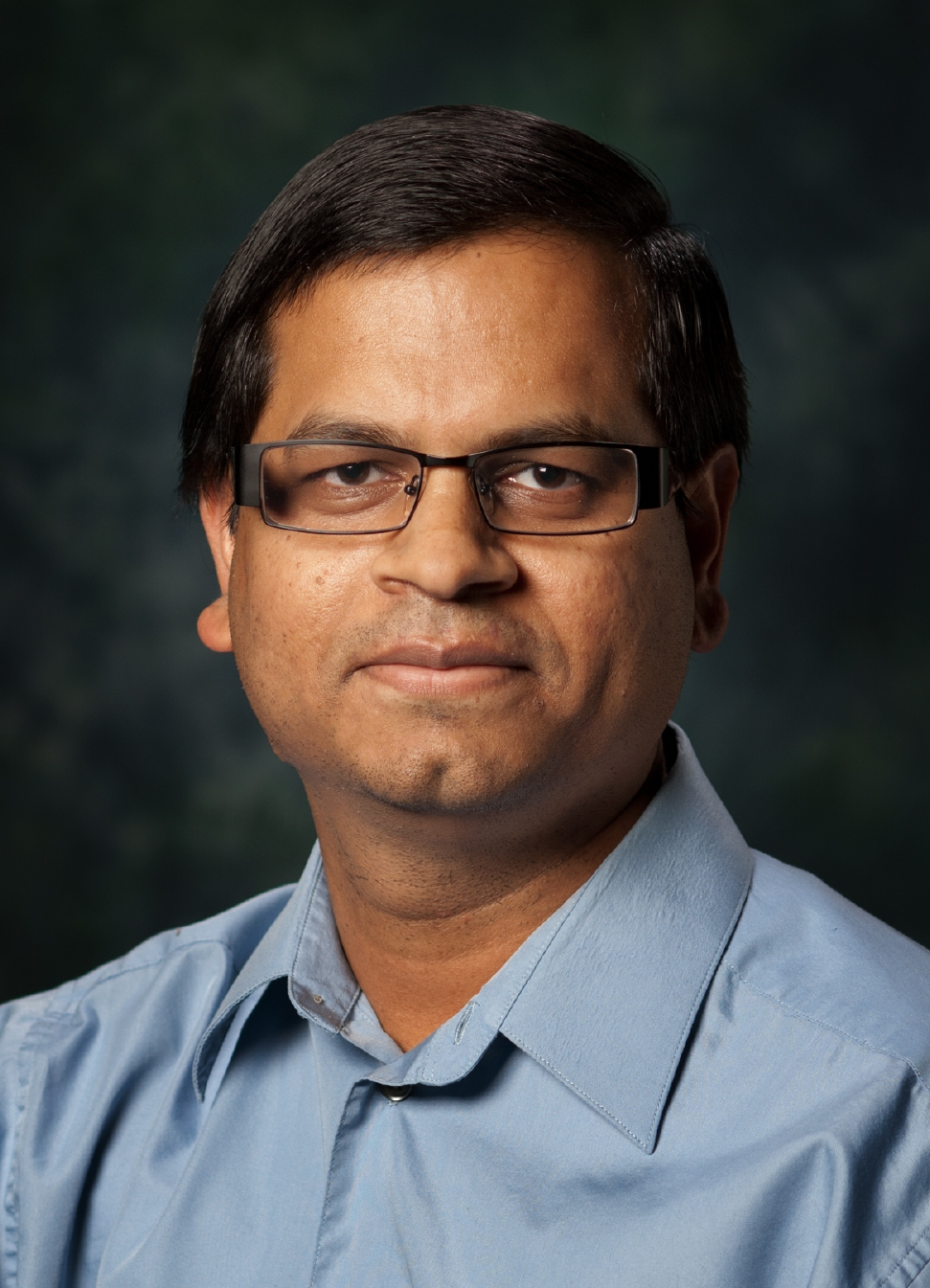}
	\vspace{-0.9cm}
\end{wrapfigure}
\textbf{Saraju P. Mohanty} obtained a Bachelors degree with Honors in Electrical Engineering from the Orissa University of Agriculture and Technology (OUAT), Bhubaneswar, 1995. His Masters degree in Systems Science and Automation is from the  the Indian Institute of Science (IISc), Bangalore, in 1999. He obtained a Ph.D. in Computer Science and Engineering (CSE) in 2003, from the University of South Florida (USF), Tampa.
He is a Professor at the University of North Texas. His research is in ``Smart Electronic Systems'' which has been funded by National Science Foundations, Semiconductor Research Corporation, US Air Force, IUSSTF, and Mission Innovation Global Alliance.
He has authored 300 research articles, 4 books, and invented 4 US patents. His Google Scholar h-index is 31 and i10-index is 110. He has received 6 best paper awards and has delivered multiple keynote talks at various International Conferences. He received IEEE-CS-TCVLSI Distinguished Leadership Award in 2018 for services to the IEEE, and to the VLSI research community.
He has been recognized as a IEEE Distinguished Lecturer by the Consumer Electronics Society during 2017-2018. 
He was conferred the Glorious India Award in 2017 for his exemplary contributions to the discipline. He received Society for Technical Communication (STC) 2017 Award of Merit for his outstanding contributions to IEEE Consumer Electronics Magazine. 
He was the recipient of 2016 PROSE Award for best Textbook in Physical Sciences \& Mathematics category from the Association of American Publishers for his Mixed-Signal System Design book published by McGraw-Hill in 2015. 
He was conferred 2016-17 UNT Toulouse Scholars Award for sustained excellent scholarship and teaching achievements. 
He is the Editor-in-Chief of the IEEE Consumer Electronics Magazine. He served as the Chair of TC on VLSI, IEEE Computer Society during 2014-2018.

\vspace{0.4cm}
%\item
\begin{wrapfigure}{l}{1.0in}
	\vspace{-0.5cm}
	\includegraphics[width=1.0in,keepaspectratio]{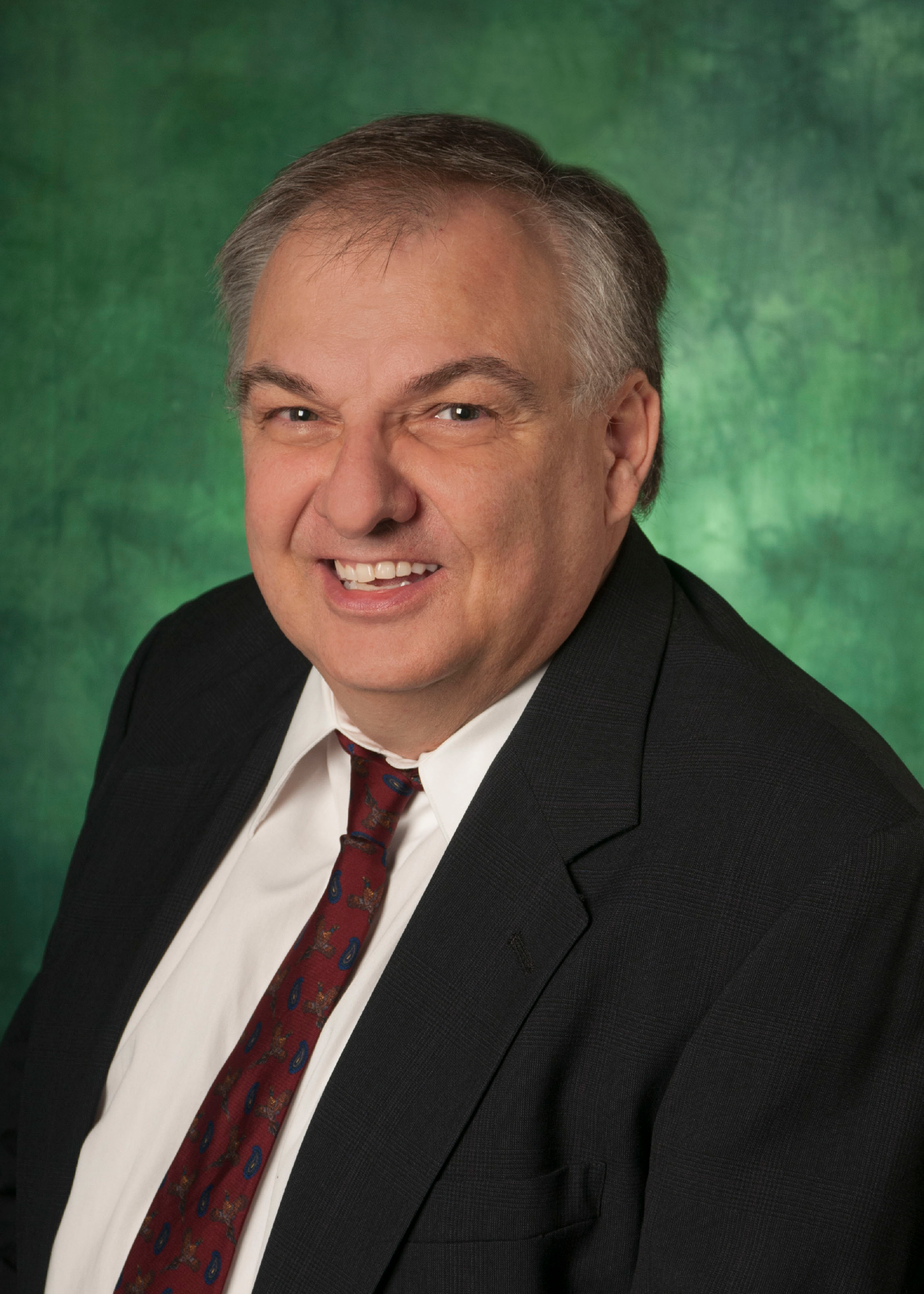}
	\vspace{-0.9cm}
\end{wrapfigure}
\noindent
\textbf{Elias Kougianos}  received a BSEE from the University of Patras, Greece in 1985 and an MSEE in 1987, an MS in Physics in 1988 and a Ph.D. in EE in 1997, all from Lousiana State University. 
From 1988 through 1998 he was with Texas Instruments, Inc., in Houston and Dallas, TX. Initially he concentrated on process integration of flash memories and later as a researcher in the areas of Technology CAD and VLSI CAD development. 
In 1998 he joined Avant! Corp. (now Synopsys) in Phoenix, AZ as a Senior Applications engineer and in 2001 he joined Cadence Design Systems, Inc., in Dallas, TX as a Senior Architect in Analog/Mixed-Signal Custom IC design. He has been at UNT since 2004. He is a Professor in the Department of Engineering Technology, at the University of North Texas (UNT), Denton, TX. His research interests are in the area of Analog/Mixed-Signal/RF IC design and simulation and in the development of VLSI architectures for multimedia applications. 
He is an author of over 120 peer-reviewed journal and conference publications. 
%\end{description}

\end{document}